\documentclass[a4paper,12pt]{article} 
\usepackage[T1]{fontenc}
\usepackage{ulem}  % \sout or strikeout text  
\usepackage{xspace}
\usepackage{subdepth}
\usepackage{times}
\usepackage{booktabs}
\usepackage{graphicx}
\usepackage{epstopdf}
\usepackage[table]{xcolor}
\newcommand{\grbb}[1]{\cellcolor[gray]{0.8}{#1}}

\usepackage{multirow}	
\usepackage{amsmath, amssymb, amscd, amsthm, etex, etoolbox, euscript, enumerate,
graphics, hhline, dsfont, multicol, multirow, tabularx}	
\usepackage{multibib}
\newcites{supp}{Appendix References}

\usepackage{lscape}
\usepackage{natbib}
		
\usepackage{float}
\usepackage[font={footnotesize,singlespacing},labelfont={bf,small},skip=0.2cm]{caption}
\floatstyle{plaintop}
\usepackage[colorlinks=true]{hyperref}
\hypersetup{linkcolor=blue} 
\hypersetup{citecolor=blue}
\hypersetup{
colorlinks=true,%
citecolor=black,%
filecolor=black,%
linkcolor=black,%
urlcolor=black
}

\def\spacingset#1{\renewcommand{\baselinestretch}%
{#1}\small\normalsize} \spacingset{1}

\makeatletter
\renewcommand*{\@fnsymbol}[1]{\ensuremath{\ifcase#1\or \dagger \or \ddagger  
\or \mathsection \or * \or \mathparagraph\or \|\or **\or \dagger\dagger
\or \ddagger\ddagger \else\@ctrerr\fi}}
\makeatother

\newtheorem{proposition}{Proposition}

\newtheorem{remark}{Remark}[proposition]

\usepackage{color}

\newcommand{\new}[1]{#1}
\newcommand{\ubar}[1]{\mkern1mu\underline{\mkern-1mu #1\mkern-1mu}\mkern1mu}

\def\bx{\mathbf x}
\def\by{\mathbf y}
\def\bX{\mathbf X}
\def\bM{\mathbf M}
\def\bz{\mathbf z}

\def\ie{\textit{i.e.}}
\def\eg{\textit{e.g.}}
\def\argmax{\text{argmax}}
\def\bbeta{\boldsymbol{\beta}}

\def\btau{\boldsymbol{\tau}}
\def\btheta{\boldsymbol{\theta}}

\def\NORM{\mathcal{N}}
\def\NIG{\mathcal{NIG}}
\def\IG{\mathcal{IG}}

\def\gpriors{\mbox{g-priors}\xspace} 

\newcommand{\1}[1]{\mathds{1}\left\{ {#1} \right\}} %indicator function

\title{Linking Frequentist and Bayesian\\
Change-Point Methods\thanks{We gratefully acknowledge the financial support of IVADO (\url{https://ivado.ca}), NSERC (grant RGPIN-2022-03767), and the Fonds de recherche du Qu\'ebec -- Soci\'et\'e et culture (grant \#193277). Preliminary versions of this paper were presented at the 2019 annual R/Finance conference in Chicago, at the Econometric Society \& Bocconi University Virtual World Congress 2020 and at the 35th European Economic Association Virtual Congress 2020. We also benefited from helpful comments from Luc Bauwens, Keven Bluteau, Kris Boudt, Linda Mhalla, Jeroen Rombouts, and Yong Song.}}
\author{David Ardia
\!\thanks{GERAD \& Department of Decision Sciences, HEC Montr\'eal, Canada}
\;\;\;\;\; Arnaud Dufays
\!\thanks{Faculty of Data Science, Economics \& Finance, EDHEC Business School, France}\,\,\thanks{\textit{Corresponding author:}
\href{mailto:arnaud.dufays@edhec.edu}{arnaud.dufays@edhec.edu}.}
\;\;\;\;\; Carlos Ord\'as Criado
\!\thanks{Department of Economics, Laval University, Canada}
\vspace{1cm}}

\date{\today}
\newcommand{\STAB}[1]{\begin{tabular}{@{}c@{}}#1\end{tabular}}

\begin{document}
\maketitle
\thispagestyle{empty}

\begin{abstract}
\noindent
We show that the two-stage minimum description length (MDL) criterion widely used to estimate linear change-point (CP) models corresponds to the marginal likelihood of a Bayesian model with a specific class of prior distributions. This allows results from the frequentist and Bayesian paradigms to be bridged together. 
Thanks to this link, one can rely on the consistency of the number and locations of the estimated CPs and the computational efficiency of frequentist methods, and obtain a probability of observing a CP at a given time, compute model posterior probabilities, and select or combine CP methods via Bayesian posteriors. Furthermore, we adapt several CP methods to take advantage of the MDL probabilistic representation. Based on simulated data, we show that the adapted CP methods can improve structural break detection compared to state-of-the-art approaches. \new{Finally, we empirically illustrate the usefulness of combining CP detection methods when dealing with long time series and forecasting.}
\end{abstract}
JEL Classification: C11, C12, C22, C32, C52, C53.\\
Keywords: change-point, model selection/combination, structural change, minimum description length.

\newpage
\clearpage
\pagenumbering{arabic} 
\newpage
\spacingset{1.8} % DON'T change the spacing!

\section{Introduction}

Change-point (CP) models are used to account for abrupt changes in an ordered sequence of observations. When a series contains structural shifts, the data can be modeled as disjoint segments of different stochastic processes. CP models typically lead to a better understanding of critical events and consistent estimates of model parameters for each segment and can also improve predictions. Due to these appealing features, CP models have attracted considerable interest in economics and natural sciences. The number and locations of the breaks are generally unknown to the researcher and must be estimated. Many methods are now available to detect CPs both in frequentist and Bayesian frameworks \citep[see,~\eg,][]{eckley2011analysis,Bauwens2013}

Frequentist and Bayesian CP approaches have mostly evolved independently in the literature, as they often explore different aspects of the CP problem. The frequentist approach has mainly focused on developing asymptotic frameworks for break detection \citep[\eg,][]{YaoAu1989,LiuAl1997, ciuperca2011general,davis2013consistency}, testing the number and location of breaks \citep[\eg,][]{BaiPerron1998}, and providing efficient algorithms to explore possible segments \citep[\eg,][]{bai1997estimating,BaiPerron2003,jackson2005algorithm,killick2012optimal,Fryzlewicz2014,YauZhao2016}. Depending on whether a CP detection method explores all possible partitions simultaneously or sequentially for a subset of segments, CP detection can be classified as global (or exact) or local, respectively. In contrast, the Bayesian CP literature has mainly focused on developing schemes for efficiently drawing from the posterior distribution of the CPs \citep[\eg,][]{carlin1992hierarchical,Stephens1994,Chib98,Kohn2008,maheu2018efficient} and on forecasting future breaks \citep[\eg,][]{PPT06,KoopPotter07}. Also, the criteria used to determine the number of CPs and their locations differ across paradigms. Frequentists mostly rely on (several variants of) the Schwarz information criterion \citep[SIC; \eg][]{YaoAu1989,LiuAl1997,ciuperca2011general,Fryzlewicz2014}, the sum of squared residuals \citep[\eg][]{bai1997estimating,BaiPerron1998}, and the two-stage minimum description length (MDL) criterion \citep[\eg][]{davis2006structural,davis2013consistency,YauZhao2016}, while Bayesians typically maximize the marginal likelihood \citep{Chib98,bernardo2007objective,Bauwens2013a,DuAl2016}. While most criteria used by frequentists consistently estimate the number and locations of the breaks, this is rarely investigated in the Bayesian framework. An exception is \citet{DuAl2016}, who establishes the consistency of the marginal likelihood to estimate the total number and location of the CPs for a piecewise-constant model.

CP model selection and combination have also been intensively studied in frequentist and Bayesian paradigms \citep[\eg,][]{ChanYipZhang2014,maheu2008learning}. Yet, links between the two statistical frameworks have been overlooked. Our paper addresses this gap by showing that the MDL criterion corresponds to the marginal likelihood of a Bayesian model with a specific class of prior distributions. Thanks to this link, the marginal likelihood inherits the asymptotic properties of the MDL criterion, and can further be used to detect CPs, assign posterior probabilities to competing CP methods, and select or combine CP methods via their posterior probabilities. Moreover, popular state-of-the-art CP methods can be adapted to the proposed marginal likelihood to improve their performance.

Similar links have been proposed in the model selection literature, in particular for normal-linear models. However, none of them tackles the CP detection problem. For example, \citet{smith1980bayes} propose deterministic functions linking the (logarithm of the) Bayes factors (ratios of marginal likelihoods) of normal-linear regression models to the Schwarz or Akaike information criteria, but their focus is on how the prior specifications impact the Bayes factor. In the same vein, \citet{KassRaftery95} emphasize that the SIC can be viewed as a rough asymptotic approximation of the (logarithm of the) Bayes factor that avoids specifying priors. \citet{HansenYu2001} recall that the MDL criterion is a first-order approximation of the marginal likelihood. They also show that the marginal likelihood can be understood as an alternative description length criterion, referred to as ``mixture MDL.''
However, these links are mainly asymptotic, the focus is on variable selection, and they ignore structural change. Our paper extends this literature to CP detection for normal-linear regression models \new{by (i) providing a class of proper priors belonging to the Normal Inverse-Gamma ($\NIG$) distributions that leads to an exact (and thus finite-sample) equivalence between the marginal likelihood of the CP model and the MDL criterion, and (ii) allowing the variance of linear models to vary over time.}

The proposed calibration has major implications for frequentist and Bayesian CP approaches. Researchers typically obtain distinct breakpoints in applied settings when different (state-of-the-art) local or global CP detection methods are used. \new{Providing a hyperparameter calibration that leads to an exact link between the marginal likelihood and a frequentist information criterion allows frequentists to rely on the Bayesian framework, make probabilistic statements about competing models, and combine or select candidate models via their posterior probabilities.} In addition, frequentists can rely on Bayesian credible intervals for the break locations rather than confidence intervals. The former are straightforward to compute and have an advantage over the latter as they do not overlap and remain in-sample. Finally, the stopping rule of several sequential detection methods, such as binary segmentation, can be adapted to reflect a probabilistic statement between competing models. This feature strongly improves the performance of sequential methods. The proposed framework also benefits Bayesians. It provides hyperparameter values that motivate the prior specification of the Bayesian CP model as they allow the marginal likelihood to inherit asymptotic properties established in the frequentist literature. Asymptotic properties for the marginal likelihood have been developed in \citet{DuAl2016} for piecewise-constant models. The proposed link allows us to extend this result to piecewise-linear Bayesian normal regression models. The consistency of the marginal likelihood directly follows from the theoretical works of \citet{davis2006structural,davis2013consistency}. \new{Additionally, our framework accounts for future breaks in long time series context. It is worth noting that Bayesian methods that allow for future breaks, such as \citet{PPT06}, \citet{KoopPotter07}, and \citet{maheu2018efficient}, cannot be applied to long time series due to their algorithmic complexity.}

In practice, global detection methods cannot always be applied due to their computational cost. Local approaches, such as binary segmentation and its extensions, are feasible alternatives \citep[\eg,][]{bai1997estimating,Fryzlewicz2014,YauZhao2016,korkas2017multiple}. Our paper broadens the set of existing CP detection methods in three ways: (i) by proposing three local algorithms that take advantage of the probabilistic representation of the MDL criterion, (ii) by adapting a fast and deterministic global method following \citet{eckley2011analysis}, and (iii) by showing how to compute credible intervals for the estimated CPs and regression parameters, which account for both in-sample and out-of-sample breaks as well as model uncertainty. These novel approaches are contrasted with existing ones using simulated data. They detect and locate breaks with the same or better accuracy than existing methods. \new{Two applications illustrate the relevance of combining local CP detection methods when dealing with long time series and forecasting. In the first application, we estimate a five-factor Fama-French CP model over a long time period by combining local CP methods. We explore the evolution of risk exposure over time while accounting for both method and break uncertainty. The large set of detected breakpoints allows us to accurately infer the break process and to provide credible intervals for risk exposures when future breakpoints occur. In the second application, we forecast the monthly U.S. inflation and demonstrate that accounting for breaks and model uncertainty leads to significant improvements in the predictive performance of the model in terms of root mean square forecast errors and mean absolute forecast errors. In particular, allowing for an out-of-sample break leads to better predictions over longer time horizons.}

The paper is organized as follows. In Section~\ref{sec:metho}, we present the methodology. Section~\ref{sec:newmethods} introduces adapted local and global CP detection methods that build upon the probabilistic representation of the MDL criterion. We test how the adapted methods collectively perform in a simulation exercise in Section~\ref{sec:mc}.  Section~\ref{sec:mcmc} proposes a full Bayesian model providing posterior distributions that account for break uncertainty and model uncertainty. \new{Section~\ref{sec:empirics} develops the two applications, and} Section~\ref{sec:conclusion} concludes.

\section{Change-Point Framework} 
\label{sec:metho}

\subsection{Model and Assumptions} 
\label{sec:cp}

Consider the following linear regression model with no structural breaks:
\begin{equation}%~\label{eq:linreg}
y_t = \beta_0 + \beta_1 x_{t,1} + \ldots + \beta_{K-1} x_{t,K-1} + \sigma \epsilon_t =  \bx_t'\bbeta + \sigma \epsilon_t \,, \quad t=1,\ldots,T \,,
\end{equation}
where $y_t$ is the observed dependent variable, $\mathbf x_t = (1, x_{t,1}, \ldots,  x_{t,K-1})'$ is a $(K\times 1)$ vector including a constant and $K-1$ covariates, $\bbeta = (\beta_0, \beta_1, \ldots, \beta_{K-1})'$ and $\sigma > 0$ are unknown parameters, and $\epsilon_t \sim iid\,\NORM(0,1)$.\footnote{While restrictive, several recent theoretical and empirical CP frameworks rely on the normality assumption (see,~\eg, \cite{rigaill2012exact,MaheuSong2013,maheu2018efficient,smith2020equity} in the Bayesian paradigm and \cite{safikhani2020joint} in the frequentist paradigm). Our calibration applies in these contexts as well.}
In applied settings, it is often unreasonable to assume that $\bbeta$ and $\sigma$ are time-invariant, especially for long time series. A more flexible linear specification allowing for $m \geq 0$ breaks ($m+1$ regimes) in the model parameters is:
\begin{equation}~\label{eq:linregbreak}
y_t  =  \bx_t'\bbeta_i + \sigma_i \epsilon_t \,, \quad \text{for}\quad \tau_{i-1} < t \leq \tau_{i} \,, \quad i=1,\ldots,m+1\,,
\end{equation}
where $\tau_i$ denotes the $i$-th CP location on the $[0,T]$ time segment, and $\bbeta_i$ and $\sigma_i$ are the model parameters holding for the period starting after time $\tau_{i-1}$ and ending at time $\tau_i$, with $\tau_{i-1} < \tau_i$. The boundary conventions $\tau_0 = 0$ and $\tau_{m+1} = T$ are used. To simplify notation, the vectors of observations belonging to segment~$i$ are denoted by $\by_i = \by_{(\tau_{i-1}+1):\tau_{i}} = (y_{\tau_{i-1}+1},\ldots,y_{\tau_i})'$ for the dependent variable and $\bX_i = [\bx_{\tau_{i-1}+1}~ \bx_{\tau_{i-1}+2}~\cdots~\bx_{\tau_{i}}]'$ for the covariates. Specification~\eqref{eq:linregbreak} encompasses stationary piecewise-linear autoregressive processes of order $q_i$ (\ie, CP-AR($q_i$)). We collect the CPs located on the $]0,T[$ interval in vector $\btau = (\tau_1,\ldots,\tau_m)'$ and gather the segment parameters in $\Theta = \{\btheta_1,\ldots,\btheta_{m+1}\}$, where $\btheta_i = (\bbeta_i,\sigma_i^2)$. The number of CPs, $\text{dim}(\btau) = m$, is typically restricted by a fixed upper bound $M \leq T$. We further assume that $\bX_i'\bX_i$ is nonsingular for all considered segments. The purpose of a CP method is to estimate the number of CPs $m$, break locations~$\btau$, and regression parameters~$\Theta$.

\subsection{Marginal Likelihood} 
\label{sec:ml}

Given the wide variety of CP methods available, model selection and model averaging are of particular interest in our setup. Model selection deals with model uncertainty by selecting one model from a set of $P$ candidate models $\{\mathcal{M}_1,\ldots,\mathcal{M}_P\}$ based on some selection criterion (\eg, AIC, SIC, MDL). Model averaging deals with model uncertainty by combining several models from a set of candidate models based on a weighting scheme. In Bayesian statistics, dealing with model uncertainty is straightforward in theory: Model selection and model averaging can be achieved by using model posterior probabilities. To compute these probabilities, the marginal likelihood function under each model needs to be evaluated. In general, this step is not trivial. However, in some specific cases, analytical expressions exist. For instance, in the context of normal-linear regression with no structural change, the marginal likelihood exhibits a closed-form expression for $\NIG$ priors \citep[see,~\eg,][]{fernandez2001benchmark}. This closed form extends to piecewise-linear models conditional on the CPs (see,~\eg, \citet{DuAl2016} and Proposition~\ref{prop:mlnig} below). 
Without loss of generality and for the sake of simplicity, we drop the conditioning on $\bX_i$ in what follows. Consider first the general expression of the marginal likelihood for the set of $m+1$ independent regimes associated with vector~$\btau$: 
\begin{equation}~\label{eq:ml}
f(\by_{1:T}|\btau)  = \prod_{i=1}^{m+1} f(\by_i | \btau) \,,
\end{equation}
where $f(\by_i | \btau)$ is the density of observing the data over the period $]\tau_{i-1},\tau_{i}]$,\footnote{When the segments are not independent, \eqref{eq:ml} can be replaced with $f(\by_{1:T}|\btau)  = \prod_{i=1}^{m+1} f(\by_{(\tau_{i-1}+1):\tau_{i}}|\btau,\by_{1:\tau_{i-1}})$, where $\by_{1:0} = \{\emptyset\}$.}  and where the model parameters $\Theta$ have been integrated out. Proposition~\ref{prop:mlnig} recalls a well-known result in Bayesian statistics and shows the prior distributions as well as the analytical expression of the marginal likelihood, which are used throughout the paper.

\new{\begin{proposition}~\label{prop:mlnig}
Consider the piecewise-linear regression model and the following $\NIG$ priors:
\begin{align} 
\begin{split}
\bbeta_i | \sigma_i^2, \btau 
& \sim \NORM\left(\ubar{\bbeta}_i,\sigma_i^2 \ubar{g}_i \ubar{\bM}_i^{-1}\right)\,, \\
\sigma_i^2 |\btau 
& \sim \IG \left( \frac{\ubar{\nu}_i}{2},\frac{\ubar{s}_i}{2} \right) \,, 
\end{split}
\end{align}
for $i = 1,\ldots,m+1$, where $\ubar{\bbeta}_i \in \mathbb{R}^K$, $\ubar{\bM}_i \in \mathbb{R}^{K \times K}$ is a definite positive matrix, $\ubar{v}_i>0$ and $\ubar{s}_i>0$ are the hyperparameters holding in segment $i$. The marginal likelihood of the observations belonging to segment $i$ is then given by:
\begin{equation} \label{eq:mlnig}
f^{\ubar{\bbeta}_i}(\by_{i}|\btau)
= (2\pi)^{-\frac{n_i}{2}} 
\left( \frac{|\bar{\bM}_i^{-1}|}{|\ubar{g}_i \ubar{\bM}_i^{-1}|} \right)^{\frac{1}{2}} \frac{\Gamma(\frac{\bar{\nu}_i}{2})}{\Gamma(\frac{\ubar{\nu}_i}{2})}    \frac{(\frac{\ubar{s}_i}{2})^{\frac{\ubar{\nu}_i}{2}}}{(\frac{\bar{s}_i}{2})^{\frac{\bar{\nu}_i}{2}}} \,,
\end{equation}
where $\bar{\nu}_i = n_i+\ubar{\nu}_i$, $\bar{s}_i = \ubar{s}_i+s_i + \ubar{\bbeta}_i'\ubar{g}_i^{-1} \ubar{\bM}_i \ubar{\bbeta}_i + \widehat{\bbeta}_i'\bX_i'\bX_i\widehat{\bbeta}_i - \bar{\bbeta}_i'\bar{\bM}_i\bar{\bbeta}_i$, $\bar{\bbeta}_i = [\ubar{g}_i^{-1} \ubar{\bM}_i + \bX_i'\bX_i]^{-1}\big(\ubar{g}_i^{-1} \ubar{\bM}_i \ubar{\bbeta}_i + \bX_i'\bX_i \widehat{\bbeta}_i\big)$ and in which $\widehat{\bbeta}_i$ and $s_i$ denote the ordinary least-squares (OLS) estimates of $\bbeta_i$ and the related sum of squared residuals computed for segment~$i$, respectively.
\end{proposition}
\begin{proof}
See the appendix, Section~\ref{app:prop:mlnig}.
\end{proof}}

Marginal likelihood $f^{\ubar{\bbeta}_i}(\by_{i}|\btau)$ in~\eqref{eq:mlnig} depends on hyperparameters $\ubar{\bbeta}_i$, $\ubar{\bM}_i$, $\ubar{v}_i$, and $\ubar{s}_i$, that we treat as constants \citep[see, \eg.,][]{DuAl2016}. \new{We further propose to set the constants so that $f^{\ubar{\bbeta}_i}(\by_{i}|\btau)$ reduces to the MDL criterion, which is commonly used in the frequentist CP literature to estimate piecewise-linear regression models (see Section~\ref{sec:mlmdl}).} With well-justified presets at hand, $f^{\ubar{\bbeta}_i}(\by_{i}|\btau)$ can be used to compute model posterior probabilities and make probabilistic statements about competing models.

To illustrate the latter point, consider the set of $P$ vectors $\{\btau^1,\ldots,\btau^P\}$ and let prior $f(\btau^p)$ be uniform. Then, the posterior probability of $\btau^p \ (p=1,\ldots,P)$ is given by:
\begin{equation}~\label{eq:probtau}
f(\btau^p | \by_{1:T}) 
=\frac{f(\by_{1:T} | \btau^p)}{\sum_{j=1}^{P}f(\by_{1:T} |\btau^j)} \,.
\end{equation}
By using \eqref{eq:ml} and \eqref{eq:mlnig} in~\eqref{eq:probtau}, the latter expression becomes feasible. Thus, we can use these model posterior probabilities for, say, selecting the candidate vector of CPs that exhibits the greatest posterior probability (\ie, $\max_{p\in \{1,P\}} f(\btau^p | \by_{1:T})$) for the given data, or for combining CP methods and averaging posterior quantities of interest by weighting these posterior quantities by the corresponding model posterior probabilities (see Section~\ref{sec:mcmc}). For instance, model uncertainty can be accounted for in the estimates of, say, parameter $\theta_t\ (t=1,\ldots,T)$, by mixing the parameter posterior densities as follows:
\begin{equation}~\label{eq:modeluncertainty}
f(\theta_t | \by_{1:T}) 
= \sum_{p=1}^{P} f(\theta_t | \by_{1:T},\btau^p)f(\btau^p | \by_{1:T}) 
= \sum_{p=1}^{P} f(\theta_t | \by_{(\tau^p_{k_p-1}+1):\tau^p_{k_p}}) f(\btau^p | \by_{1:T}) \,,
\end{equation}
where $t \in \ ] \tau^p_{k_p-1}, \tau^p_{k_p}]$ and index $k_p$ denotes the $k$\textsuperscript{th} segment of the $p$\textsuperscript{th} model ($k_p=0,\ldots,m_p+1$). These model selection and model combination approaches are illustrated in Sections~\ref{sec:mc} and~\ref{sec:empirics}. In the context of forecasting, the $h$-step-ahead $(h \geq 1)$ predictive density would similarly account for model uncertainty through the weighted average:
\begin{equation}
f(\by_{T+1:T+h}|\by_{1:T}) 
= \sum_{p=1}^{P} f(\by_{T+1:T+h}|\by_{(\tau^p_{k_p-1}+1):\tau^p_{k_p}})f(\btau^p|\by_{1:T}) \,.
\end{equation}
Section~\ref{sec:empirics} explores forecasting performance accounting for model uncertainty.

\subsection{MDL Marginal Likelihood}
\label{sec:mlmdl}

As emphasized earlier, the MDL criterion is widely used to (consistently) estimate piecewise-stationary autoregressive models \citep[AR;~\eg,][]{davis2006structural,killick2012optimal,YauZhao2016,ng2022bootstrap}. In this section, we build upon Proposition~\ref{prop:mlnig} and set the hyperparameters of the $\NIG$ priors so that the marginal likelihood reduces to the MDL criterion. 

For a piecewise-linear model with $m$ breaks, the MDL criterion reads:
\begin{equation}~\label{eq:mdl}
\text{MDL}(m,\btau) = \ln f(\by_{1:T}|\btau,\widehat \Theta_{\text{MLE}}) 
-\ln^{\!+}\!(m) - (m+1)\ln T - \left( \frac{K+1}{2} \right) \sum_{i=1}^{m+1}\ln n_i \,,
\end{equation}
where $f(\by_{1:T}|\btau,\widehat \Theta_{\mathrm{MLE}})$ is the conditional likelihood function, $\widehat \Theta_{\mathrm{MLE}}$ is the set of model parameters maximizing the likelihood for the given vector $\btau$, and $\ln^{\!+}(x) = \max\left\{0,\ln(x)\right\}$ for $x \geq 0$. Proposition~\ref{prop:mlmdl} formally establishes the link between a class of Bayesian CP models and the MDL criterion.

\new{\begin{proposition}~\label{prop:mlmdl}
Under the setting in Proposition~\ref{prop:mlnig}, let the hyperparameters of the $\NIG$ priors be given by:
\begin{equation} \label{eq:mlmdl1}
\ubar{\bbeta}_i = \widehat{\bbeta}_i \,,\,\, 
\ubar{\bM}_i = \bX_i'\bX_i \,,\,\, 
\ubar{g}_i = \ubar{f}_{\!i} n_i  - 1\,,\,\,
\ubar{f}_{\!i} = \left(\frac{((m^{+})^{\frac{1}{m+1}})n_i^{\frac{1}{4}}T}{(\frac{1}{\sqrt{n_i}} +1)^{\frac{1}{2}}}\right)^{\frac{2}{K}} 
\!\!\!\! \exp \left( \frac{2}{K}\Delta R_{4,i} \right)\,,\,\,
\end{equation}
\begin{equation} \label{eq:mlmdl2}
\ubar{\nu}_i = \ubar{k}_i n_i \,,\,\,
\ubar{s}_i = \ubar{k}_i s_i\,,\,\,
\ubar{k}_i = \frac{1}{\sqrt{n_i}}\,,\,
\end{equation}
for $i = 1,\ldots,m+1$, where $m^{+}=\max\left\{1,m\right\}$, $\Delta R_{4,i} = R_4(\frac{n_i+\ubar{\nu}_i}{2}) - R_4(\frac{\ubar{\nu}_i}{2})$ with $R_4(x)= \frac{1}{12 x} - \frac{1}{360 x^3} + \frac{1}{1260 x^5}$ for $x>0$. Then, the logarithm of the marginal likelihood conditional on $\btau$ is given by:
\begin{equation}~\label{eq:mlmdl}
\ln f^{\mathrm{MDL}}(\by_{1:T} | \btau) 
=\sum_{i=1}^{m+1}\ln f^{\widehat{\bbeta}_i}(\by_i|\btau) = \mathrm{MDL}(m,\btau) + \mathcal{O}\left(\min_{i} (n_i)^{-\frac{7}{2}}\right) \,,
\end{equation}
where $\min_{i} (n_i)$ denotes the number of observations in the smallest regime.
\end{proposition}
\begin{proof}
See the appendix, Section~\ref{app:prop:mlmdl}.
\end{proof}
\begin{remark}~\label{rem:mle}
The hyperparameters of the inverse-gamma distribution imply that the maximum likelihood estimator $\frac{s_i}{n_i}$ lies in a high-density region of the prior as it falls between the 
mode and the expectation; see the appendix, Section~\ref{app:rem:mle}.
\end{remark}
\begin{remark}~\label{rem:approx}
The approximation order in \eqref{eq:mlmdl} can be made arbitrarily small by selecting a high-order $N$ in the remainder $\Delta R_{N,i}$; see the appendix, Section~\ref{app:prop:mlmdl}.
\end{remark}}

In the rest of the paper, we refer to the marginal likelihood obtained in Proposition~\ref{prop:mlmdl} as the \textit{MDL marginal likelihood}. Given that $\ubar{\bM}_i = \bX_i'\bX_i$ in~\eqref{eq:mlmdl1}, the priors suggested in Proposition~\ref{prop:mlmdl} belong to the class of \gpriors \citep[see,~\eg,][]{zellner1986assessing,fernandez2001benchmark}. 

Besides being simple to use, the calibration proposed in Propositions~\ref{prop:mlmdl} is of great interest to link the frequentist and Bayesian statistical paradigms. For example, the $\NIG$-prior distributions used in the proposition are also used in many Bayesian CP papers to improve the estimation efficiency \citep[see, \eg,][]{Geweke2011,rigaill2012exact,MaheuSong2013}. Our calibrations further extend the consistency properties of the MDL criterion to these Bayesian CP setups. The proposed prior hyperparameters lead to an exact equivalence between the marginal likelihood and the MDL criterion. \citet[][]{davis2013consistency} show that the MDL criterion is a consistent estimator of the number of CPs, their locations, the order of the linear model, and the parameter values in each segment for a broad class of piecewise-linear stationary autoregressive processes, which nest canonical model~\eqref{eq:linregbreak}. These asymptotic properties also apply to the MDL marginal likelihood under the regularity conditions stated in \citet{davis2013consistency} and under the presets given in Proposition~\ref{prop:mlmdl}. Moreover, by establishing the equivalence between the MDL criterion and the marginal likelihood for a specific class of priors, we extend the consistent maximum likelihood estimator of \citet{DuAl2016} from a piecewise-constant to a piecewise-linear multiple regression setup. The consistency of the MDL marginal likelihood is shown numerically in Table~\ref{tab:simulation:large} of the simulation section.

\section{Revisiting CP Methods}
\label{sec:newmethods}

In this section, we revisit three popular CP methods (two local ones based on binary segmentation and the efficient and global method of \citet{BaiPerron2003}) by adapting them to the MDL marginal likelihood criterion. In addition, we propose a fourth method that ``prunes'' the adapted global approach and reduces its complexity to an order lower than $\mathcal{O}(T^2)$. The simulation results reported in Section~\ref{sec:mc} demonstrate that the pruning achieves a good compromise between accuracy in CP detection and computational cost, particularly for long time series. 

\subsection{Binary and Wild Binary Segmentation} 
\label{ss:bs}

Binary segmentation, denoted by BS or $\text{BS}(a,b)$ hereafter, is a generic procedure that sequentially detects multiple breaks in time series over a finite time segment $[a,b]$ with $a<b$ \citep[see, \eg, ][]{gupta1996detecting}. BS relies on the ``cumsum'' statistic:
\begin{equation}~\label{eq:bs}
\widetilde{y}_{a,b}(\tau) 
= \sqrt{\frac{b-\tau}{(b-a+1)(\tau-a+1)}} \sum_{t=a}^{\tau}y_t - \sqrt{\frac{\tau-a+1}{(b-a+1)(b-\tau)}} \sum_{t=\tau+1}^{b}y_t \,,
\end{equation}
where  $a< \tau < b$. When searching over the entire time span $[1,T]$, $\text{BS}(a,b)$ starts by setting $a=1$ and $b=T$ and then operates as follows:
\begin{itemize}
\item for $\tau \in (a,b)$, compute $\overline{\tau} = \argmax_{\tau} \widetilde{y}_{a,b}(\tau)$;
\item  if $\widetilde{y}_{a,b}(\overline{\tau})>\delta_T$, add $\overline{\tau}$ to the set of CPs. Then, run $\text{BS}(a,\overline{\tau})$ and $\text{BS}(\overline{\tau},b)$. 
\end{itemize}
If none of the CP candidates yield a statistic $\widetilde{y}_{1,T}(\tau)$ strictly above $\delta_T$, the algorithm stops, and the series does not exhibit structural change. Threshold $\delta_T$ is chosen based on theoretical and empirical considerations. As noted by \citet[][]{Fryzlewicz2014}, if the series exhibits similarities across regimes (such as short regime duration or offsetting jumps at adjacent segments), BS can lead to a test statistic that is below the threshold. To overcome this weakness, this author proposes an alternative procedure called ``wild binary segmentation'' and denoted as WBS or $\text{WBS}(a,b)$ hereafter. When applied to the observations belonging to the entire interval $[1,T]$, the algorithm starts by setting $a=1$ and $b=T$ and proceeds as follows:
\begin{itemize}
\item generate a set of $N$ random intervals $\{a_n,b_n\}_{n=1}^{N}$ within $[a,b]$. 
\item for each interval, apply $\text{BS}(a_n,b_n)$ and save the maximum value of the statistic\\$\widetilde{y}_{a,b}(\overline{\tau}) = \text{max}_{n \in [1,N]}\widetilde{y}_{a_n,b_n}(\overline{\tau}_n)$, where $\overline{\tau}_n = \argmax_\tau \widetilde{y}_{a_n,b_n}(\tau)$.
\item if $\widetilde{y}_{a,b}(\overline{\tau})>\delta_T$, set $\overline{\tau}$ as a new CP. Then, run $\text{WBS}(a,\overline{\tau})$ and $\text{WBS}(\overline{\tau},b)$.
\end{itemize}

Both BS and WBS rely on fitting criterion~\eqref{eq:bs}. By replacing this criterion with:
\begin{equation}~\label{eq:bayesianbs}
\widetilde{y}_{a,b}^{\mathrm{MDL}}(\tau) 
= \ln \frac{f^{\mathrm{MDL}}(\by_{a:b}|\btau=\{a-1,\tau,b\})}{f^{\mathrm{MDL}}(\by_{a:b}|\btau=\{a-1,b\})},\,
\end{equation}
threshold $\delta_T$ can be given a probabilistic justification. That is, when $\widetilde{y}_{a,b}^{\mathrm{MDL}}(\tau) > \delta_T = 0$, the posterior probability of a model exhibiting a break at location $\tau$ on interval $[a,b]$ is greater than that of its no-break counterpart. When $\delta_T > 3$, the probability in favor of a break is greater than 95\%. By setting $\delta_T = 3$, we can readily adapt BS and WBS to the Bayesian criterion~\eqref{eq:bayesianbs}. 
We call these two modified CP methods based on the MDL marginal likelihood ``BSMDL'' and ``WBSMDL,'' respectively, and test their performance in the simulation section.\footnote{Note that this adaptation is proposed for the BS algorithm in the context of CP Poisson processes by \citet{young2001bayesian}. In that case, however, the threshold $\delta_T$ is set to $0$, and the prior for $\tau$ is a noninformative (uniform) distribution.} 

\subsection{\citet{BaiPerron2003}'s Approach}
\label{sec:bp}

The efficient global algorithm proposed by \citet{BaiPerron2003} is one of the most popular algorithms for detecting multiple CPs in linear regression models. For a bounded set of breaks $m=1,\ldots, M$, this algorithm relies on dynamic programming to determine the CP locations; it minimizes the sum of squared residuals for each number of breakpoints in the set. The procedure requires at most least-squares operations of $\mathcal{O}(T^2)$, which is far fewer than the brute force approach of $\mathcal{O}(T^{m})$ for any $m>2$. It further handles minimum regime duration and can readily be used with a specific class of information criteria.

A CP information criterion can be optimized with this algorithm if it can be decomposed into the following additive components:
\begin{equation}~\label{eq:condbp}
\text{IC}(m,\btau,\by_{1:T}) = u(m,T) + \sum_{i=1}^{m+1} v(\by_i|\tau_{i-1},\tau_i) \,,
\end{equation}
where $u(m,T)$ is a function that only depends on the number of CPs and on the sample size, and where $v(\by_i|\tau_{i-1},\tau_i)$ is a function that depends on the starting and terminal dates $\tau_{i-1}$ and~$\tau_i$. For instance, \citet{rigaill2012exact}, \citet{MaheuSong2013} and \citet{DuAl2016} use different marginal likelihood criteria based on different $\NIG$ priors that comply with~\eqref{eq:condbp}. 

We adapt this global method by replacing the sum of squared residuals with the MDL marginal likelihood. Given $m$ CPs, MDL criterion~\eqref{eq:mdl} verifies~\eqref{eq:condbp}. Indeed, as \citet{eckley2011analysis} point out, no heuristic algorithms such as the genetic ones proposed in \citet{davis2006structural} or \cite{li2012multiple} are necessary for finding the segments that maximize the MDL criterion. We call this global CP method ``GMDL'' and explore its performance in the simulation section. When multiple $m$ values are explored, one can readily use posterior probabilities~\eqref{eq:probtau} to assess the most likely number of breaks.

Global segmentation based on efficient algorithms of order $\mathcal{O}(T^2)$ can be computationally demanding for long time series. This has motivated researchers to develop local alternatives that exhibit less complexity than $\mathcal{O}(T^2)$. Below, we propose a pruned global method of complexity $\mathcal{O}(T)$. As shown in \citet{BaiPerron2003}, the total number of possible segments in their approach is at most $\frac{T(T+1)}{2}$. Figure~\ref{tab:wildglobal} illustrates this statement for a sample of size $T=10$ with a minimal duration of one time period before a potential break occurs.

To reduce the complexity of this global method, we only need to decrease the number of considered segments in Figure~\ref{tab:wildglobal} by a function proportional to $T$. In the spirit of WBS, we can evaluate $\alpha T$ randomly-selected segments, where $\alpha$ is a fixed constant. 

%\insertfloat{Figure~\ref{tab:wildglobal}}
\begin{figure}[H]
\centering
\spacingset{1} % DON'T change the spacing!
\scalebox{0.75}{
\begin{tabular}{lccccccccccc}
\toprule
&& \multicolumn{10}{c}{Terminal date} \\								
\cmidrule(lr){3-12}
&& 1 &   2 &     3 &     4 &     5 &     6 &     7 &     8 &     9 &     10\\ 
\midrule
\multirow{10}{*}{\STAB{\rotatebox[origin=c]{90}{Starting date}}}  
&1& O & 	O & 	X & 	X & 	X & 	X & 	X & 	X & 	X & 	X \\
&2&  & 	O & 	O & 	X & 	X & 	X & 	X & 	X & 	X & 	X \\ 
&3&  & 	 & 	O & 	O & 	X & 	X & 	X & 	X & 	X & 	X \\ 
&4&  & 	 & 	 & 	O & 	O & 	X & 	X & 	X & 	X & 	X \\ 
&5&  & 	 & 	 & 	 & 	O & 	O & 	X & 	X & 	X & 	X \\ 
&6&  & 	 & 	 & 	 & 	 & 	O & 	O & 	X & 	X & 	X \\ 
&7&  & 	 & 	 & 	 & 	 & 	 & 	O & 	O & 	X & 	X \\ 
&8&  & 	 & 	 & 	 & 	 & 	  & 	 & 	O & 	O & 	X \\ 
&9&  & 	 & 	 & 	 & 	 & 	  & 	 & 	 & 	O & 	O \\ 
&10&  & 	 & 	 & 	 & 	 & 	  & 	 & 	 & 	 & 	O \\
\bottomrule
\end{tabular}}
\caption{\textbf{Global method -- Number of possible segments}\\ 
The vertical number indicates the initial date of a segment and the horizontal number indicates the ending date. We set the sample size $T=10$ with a minimum regime duration of one observation (since the start) and an unspecified number of breaks $m$. For instance, in the first row, possible segments are 1 to 3, 1 to 4, ..., 1 to 10, where the initial and in-between-break dates of the possible segments are denoted by O, and the ending dates are denoted by Xs. In the fourth row, possible segments are 4 to 6, 4 to 7,..., 4 to 10. Setting a fixed value for $m$ would imply further restrictions (or O terms); see \citet[][Fig. 1]{BaiPerron2003}.}
\label{tab:wildglobal}
\end{figure}

Alternatively, following the LRSM method of \citet{YauZhao2016}, we first estimate a set of potential CPs, $\mathcal{J}^{*} = \{\tau_1^{*},\ldots,\tau_{m^*}^{*}\}$, with a local method and then find the best subset of CPs via a global method, such that:
\begin{equation}
\left(m^{**}, \mathcal{J}^{**}\right) = \argmax_{m \leq m^*\!, \, \mathcal{J} \ \subseteq \ \mathcal{J}^{*}} \text{MDL}(m, \mathcal{J}) \,.
\end{equation}
This two-step approach is now standard in the literature \citep[see][]{ChanYipZhang2014,Fryzlewicz2014}. We replace the MDL criterion of the scanning window proposed in the original LRSM approach with criterion~\eqref{eq:bayesianbs} and compute $\mathcal{J}^{*}$ as follows:  
\begin{equation}~\label{eq:mdlset}
\mathcal{J}^{*} = \left\{ l \in \{h,h+1,\ldots,T-h\}: \widetilde{y}_{1,T}^{\text{MDL}}(l) = \max_{t\in (l-h,l+h)} \widetilde{y}_{1,T}^{\text{MDL}}(t) \right\} \,,
\end{equation}
where $h$ is the closest integer to $\ln T$ and $t\in (l-h,l+h)$ is the scanning window centered on $l$ and browsing the entire $[1,T]$ segment. Since radius $h$ is a function of $T$, the complexity of the method is at most $\mathcal{O}\left((\frac{T}{\ln T})^2\right)$. While this improvement with respect to $\mathcal{O}\left(T^2\right)$ seems small, the number of segments is drastically reduced in applied settings. For instance, for each simulated series of the six DGPs considered in the simulation section, the number of potential breaks never exceeds 8\% of the sample size. Figure~\ref{tab:wildglobal2} illustrates the pruned version of the global method for a sample size of ten observations and potential locally estimated break at periods $\mathcal{J}^{*} = \{4,8\}$. We call this pruned version of the GMDL method ``PGMDL'' and test its performance in the simulation section. 

\begin{figure}[H]
\centering
\spacingset{1} % DON'T change the spacing!
\scalebox{0.75}{
\begin{tabular}{lccccccccccc}
\toprule
&& \multicolumn{10}{c}{Terminal date} \\								
\cmidrule(lr){3-12}
&& 1 &   2 &     3 &     4 &     5 &     6 &     7 &     8 &     9 &     10\\ 
\midrule
\multirow{10}{*}{\STAB{\rotatebox[origin=c]{90}{Starting date}}}   
& 1 & O & 	O & 	O & 	X & 	O & 	O & 	O & 	X & 	O & 	X \\
& 2 &  & 	O & 	O & 	X & 	O & 	O & 	O & 	X & 	O & 	X \\ 
& 3 &  & 	 & 	O & 	O & 	O & 	O & 	O & 	X & 	O & 	X \\ 
& 4 &  & 	 & 	 & 	O & 	O & 	O & 	O & 	X & 	O & 	X \\ 
& 5 &  & 	 & 	 & 	 & 	O & 	O & 	O & 	X & 	O & 	X \\ 
& 6 &  & 	 & 	 & 	 & 	 & 	O & 	O & 	X & 	O & 	X \\ 
& 7 &  & 	 & 	 & 	 & 	 & 	 & 	O & 	O & 	O & 	X \\ 
& 8 &  & 	 & 	 & 	 & 	 & 	 & 	 & 	O & 	O & 	X \\ 
& 9 &  & 	 & 	 & 	 & 	 & 	 & 	 & 	 & 	O & 	O \\ 
& 10 &  & 	 & 	 & 	 & 	 & 	 & 	 & 	 & 	 & 	O \\
\bottomrule
\end{tabular}}
\caption{\textbf{Pruned global method -- Number of possible segments}\\
The vertical number indicates the initial date of a segment and the horizontal number indicates the ending date. We set the sample size to $T=10$ with a possible set of locally estimated breaks at periods $\mathcal{J}^{*} = \{4,8\}$, and a maximal number of breaks set to $M=m^*=2$. For instance, in the first row, possible segments are 1 to 4, 1 to 8 and 1 to 10, where the initial and in-between-break dates of the possible segments are denoted by O while the terminal dates are denoted by Xs. In the fourth row, possible segments are 4 to 8 and 4 to 10.}
\label{tab:wildglobal2}
\end{figure}

\section{Simulation Study}
\label{sec:mc}

We assess the performance of the CP methods proposed in Section~\ref{sec:newmethods} in a simulation study. The goal is to: (i) compare the performance of the adapted CP methods proposed in Section~\ref{sec:newmethods} with existing ones, (ii) illustrate how we can capture the uncertainty related to the choice of a CP method and the number of breaks for inference, and (iii) show the computational gain of combining local methods for detecting breakpoints in long time series. The following CP methods are considered:

\begin{enumerate}
\item BS: Binary segmentation, a local procedure implemented in the \textsf{R} package \textbf{wbsts} \citep{korkas2017multiple,wbsts}. We use function \texttt{wbs.lsw(...,M=1)} for detecting the breakpoints, keeping the default values for all other function arguments.
\item WBS: Wild binary segmentation, a local procedure presented in \citet{korkas2017multiple} and also implemented in the R package \textbf{wbsts}. Again, function \texttt{wbs.lsw(...,M=0)} is used, keeping the default values for all other function arguments.
\item LRSM: The likelihood ratio scan method, a local procedure proposed by  \citet{YauZhao2016} and available at \url{http://wileyonlinelibrary.com/journal/rss-datasets}. We set the window size to $h = \max\{50, 2\log(T)^2\}$, the maximum AR order to 10, the minimum distance between two CPs to 50, and the minimum distance between the relative position of two CPs to 1\%.
\item BSMDL: The binary segmentation procedure based on break detection criterion~\eqref{eq:bayesianbs}. Threshold~$\delta_T$ is set to 3.
\item WBSMDL: The wild binary segmentation procedure based on break detection criterion~\eqref{eq:bayesianbs}. Again,  threshold~$\delta_T$ is set to 3.
\item PGMDL: The pruned global MDL method based on \eqref{eq:mdlset}. We use \citet{BaiPerron2003}'s algorithm to optimize the MDL criterion given the set of potential breakpoints obtained with scan statistics~\eqref{eq:mdlset}. This is also a local method. The maximum number of breaks and the minimum regime duration are set to 50 and 10$K$, respectively.
\item GMDL: The global MDL method as described under~\eqref{eq:condbp}. This corresponds to \citet{BaiPerron2003}'s approach applied to the MDL criterion for a given number $m$ of breaks. The upper bound of $m$ and the minimum regime duration are set to 50 and 10$K$, respectively.
\item SEL: This selection scheme chooses the method among local methods 1 to 6 that delivers the highest posterior probability according to~\eqref{eq:probtau}. 
\end{enumerate}

The simulated data considered are the first six stationary AR linear processes of \citet[Section~4]{YauZhao2016}, which are reproduced in Table~\ref{tab:dgp}. All DGPs are of autoregressive order up to two. DGP~A is a stationary AR(1) model with no structural change. DGPs~B and C are two standard piecewise-stationary AR processes with two breaks in the conditional mean occurring far from the start and end periods, and with constant variance. DGP~D is a single-CP model with an early break (after 50 observations) in mean and constant variance. The last two DGPs, E and F, are strongly persistent stationary piecewise-linear AR processes with changing means and variances, for which structural change is typically more difficult to detect: the near-unit-root AR parameter of DGP E is constant across segments, while the autocorrelation function of DGP F does not change much across segments.

\begin{table}[H]
\centering
\small
\spacingset{1} % DON'T change the spacing!
\caption{\textbf{Data generating processes}\\
This table reports the DGPs investigated in \citet[Section~4]{YauZhao2016} and used in the simulation study.}
\scalebox{0.70}{ 	 	 
\begin{tabular}{@{}ccccc@{}c@{}c l @{}}	 	 	 	 
\toprule
\multicolumn{8}{c}{DGP: $y_{t} = \beta_0 + \beta_{1} y_{t-1} + \beta_2 y_{t-2} + \epsilon_t$ where $\epsilon_t \sim iid\,\mathcal{N}(0,\sigma^2)$} \\
\midrule
\multirow{2}{*}{DGP} & \multirow{2}{*}{\parbox{0.8cm}{\hspace{0.2cm} \# \\ breaks}} &  \multirow{2}{*}{\parbox{1.2cm}{Break \\ location}} & \multirow{2}{*}{$\beta_0$} & \multirow{2}{*}{$\beta_1$} & \multirow{2}{*}{$\beta_2$} & \multirow{2}{*}{$\sigma^2$} & \multirow{2}{*}{Description}\\
& & \\
\midrule
A   & 0 & \{-\}  & \{0\}  & \{-0.7\}   & \{0\} & \{1\} & Stationary AR(1), no break\\
B   & 2 & \{514,768\}  & \{0,0,0\}  & \{0.9,1.69,1.32\}   & \{0,-0.81,-0.81\} & \{1,1,1\} & Piecewise-stationary AR(2), two breaks\\
C   & 2 & \{400,612\}  & \{0,0,0\}  & \{0.4,-0.6,0.5\}   & \{0,0,0\} & \{1,1,1\} & Piecewise-stationary AR(1), two breaks\\
D   & 1 & \{50\}  & \{0,0\}  & \{0.75,-0.5\}   & \{0,0\} & \{1,1\} & Piecewise-stationary AR(1), one early break\\
E   & 2 & \{400,750\}  & \{0,0,0\}  & \{0.999,0.999,0.999\}   & \{0,0,0\} & \{1,2.25,1\} & Piecewise-near-unit-root AR(1), two breaks\\
F   & 2 & \{400,750\}  & \{0,0,0\}  & \{1.399,0.999,0.699\}   & \{-0.4,0,0.3\} & \{1,2.25,1\} & Piecewise-near-unit-root AR(2), two breaks\\
\bottomrule
\end{tabular}}
\label{tab:dgp}
\end{table}

We perform 1,000 replications of the above DGPs and consider two metrics to evaluate the performance of the CP methods: (i) the proportion of replications for which a method identifies the true number of breaks, and (ii) the ``exact frequency'' defined as the proportion of replications for which a method achieves both the correctly estimated number of breaks and the absolute distance between each pair of true and estimated break locations within~50 \citep[see][Section~4.2]{YauZhao2016}.

The performance of the eight CP methods for the six DGPs and samples of size $T=1,\!024$ is reported in Table~\ref{tab:simulation}. Overall, the methods are reliable, both in detecting the true number of breaks (in bold) and their locations: 87.5\% of the reported experiments (42 out of 48) detect the true number of breaks more than 70\% of the time, and 79.2\% of them (38 out of 48) achieve that same level of accuracy in terms of exact frequency. BS and WBS clearly exhibit the worst overall performance. Note, however, that all local methods except BS and WBS detect the CPs assuming the true lag orders of the autoregressive processes.\footnote{Recall that all methods based on the MDL marginal likelihood require OLS estimates ---the $\widehat{\bbeta}_i$s defined in~\eqref{eq:mlmdl1} and in Proposition~\ref{prop:mlmdl}--- for all relevant segments as an input. For example, criterion~\eqref{eq:bayesianbs} used to compute BSMDL and WBSMDL makes use of the true lag orders of the DGPs in these OLS estimates. In comparison, the CPs obtained via BS or WBS rely on criterion~\eqref{eq:bs}, which does not require any particular DGP.} Given the latter, the results for BS and WBS are reasonably good and rather remarkable in the context of the near-unit-root processes~E and~F. In particular, BS and WBS beat the local methods LRSM and PGMDL in detecting and locating the breaks of process~E.\footnote{Note that our LRSM simulation results are more favorable for DGP E and F than those reported in~\citet[Section~4]{YauZhao2016}. These authors rely on Yule-Walker instead of maximum likelihood estimators and omit the term $\ln^{\!+}\!(m)$ so that the optimal partitioning algorithm of \citet{jackson2005algorithm} can be used.} \new{As well-known, structural breaks happening at the beginning or end of the sample are typically more difficult to detect with standard CP methods \citep[see, \eg,][]{HorvthEtAl2020}. In our simulation results, the adapted global and local methods accurately detect and locate DGP~D's early break. In particular, the performance of the SEL approach stresses the potential benefits of combining the local methods to detect early or late breakpoints.} Note also that the local methods based on the MDL marginal likelihood compare favorably with other existing state-of-the-art methods \citep[for example, see][Table~3]{YauZhao2016}. Notice that WBSMDL outperforms all methods investigated in Table~\ref{tab:simulation} (including the global method GMDL) for both performance metrics in all investigated DGPs. As expected, the global GMDL method is very reliable. For this particular Monte Carlo exercise, the SEL method performs equally well as the GMDL method.

\begin{table}[H]
\centering
\spacingset{1} % DON'T change the spacing!
\caption{\textbf{Results of the simulation study -- Performance of the CP methods}\\
This table reports the results of the simulation study for the eight CP methods (six local, one global, and one selection method) for the DGPs of Table~\ref{tab:dgp}. All simulated series are of size $T=1,\!024$. Each series has been generated 1,000 times. For each run, the SEL method picks the local CP method with the highest posterior probability of the six local methods. The ``$m=\cdot$'' columns report the proportion of total replications for which a method estimates $m$ breaks. Bold values highlight the true number of breaks in the corresponding DGP. The ``Exact'' column reports the proportion of total replications for which a method infers the true number of breaks and achieves an absolute distance between each pair of true and estimated break within~50.}
\scalebox{0.70}{
\begin{tabular}{c ccccc ccccc}
\toprule
&    & \multicolumn{4}{c}{\# of detected breaks} & 	 &   \multicolumn{4}{c}{\# of detected breaks} \\									
\cmidrule(lr){3-6}\cmidrule(lr){8-11}												
DGP & 	   	 Exact &  	 
$m\!=\!0$ & 	 $m\!=\!1$ & 	 $m\!=\!2$ & 	 $m\!\geq\!3$ 
& 	  	 Exact & 	 
$m\!=\!0$ & 	 $m\!=\!1$ & 	 $m\!=\!2$ & 	 $m\!\geq\!3$ \\									
\midrule											
	 & \multicolumn{5}{c}{BS  method } & 	  \multicolumn{5}{c}{WBS method } \\ 									
\cmidrule(lr){1-1}\cmidrule(lr){2-6}\cmidrule(lr){7-11}												
A & 	48.3  & 	\textbf{48.3}  & 	20.8  & 	19.8  & 	11.1  & 		78.8  & 	\textbf{78.8}  & 	16.8  & 	3.9  & 	0.5  \\
B & 	33  & 	0  & 	14.1  & 	\textbf{51.6}  & 	34.3  & 		27.5  & 	0  & 	29.1  & 	\textbf{53.6}  & 	17.3  \\
C & 	83.8  & 	0  & 	0.1  & 	\textbf{88.6}  & 	11.3  & 		84.7  & 	0.4  & 	0  & 	\textbf{92.3}  & 	7.3  \\
D & 	58.8  & 	12.9  & 	\textbf{66}  & 	13.6  & 	7.5  & 		69.7  & 	21.2  & 	\textbf{73.7}  & 	4.9  & 	0.2  \\
E & 	66.9  & 	0.3  & 	3.9  & 	\textbf{90.6}  & 	5.2  & 		73.6  & 	5  & 	1.1  & 	\textbf{89.9}  & 	4  \\
F & 	61.8  & 	0.2  & 	10.4  & 	\textbf{82.2}  & 	7.2  & 		73.3  & 	0  & 	5.8  & 	\textbf{89.3}  & 	4.9  \\[0.2cm]
	 & \multicolumn{5}{c}{BSMDL method } & 	  \multicolumn{5}{c}{WBSMDL method } \\ 									
\cmidrule(lr){1-1}\cmidrule(lr){2-6}\cmidrule(lr){7-11}												
A & 	100  & 	\textbf{100}  & 	0  & 	0  & 	0  & 		100  & 	\textbf{100}  & 	0  & 	0  & 	0  \\
B & 	98.5  & 	0  & 	0  & 	\textbf{100}  & 	0  & 		99.7  & 	0  & 	0  & 	\textbf{100}  & 	0  \\
C & 	99.7  & 	0  & 	0  & 	\textbf{100}  & 	0  & 		100  & 	0  & 	0  & 	\textbf{100}  & 	0  \\
D & 	71.2  & 	2.9  & 	\textbf{97.1}  & 	0  & 	0  & 		100  & 	0  & 	\textbf{100}  & 	0  & 	0  \\
E & 	69.6  & 	7.3  & 	4.2  & 	\textbf{86.5}  & 	2  & 		84.6  & 	3.4  & 	1.8  & 	\textbf{93.4}  & 	1.4  \\
F & 	79.2  & 	0  & 	3.6  & 	\textbf{94.9}  & 	1.5  & 		92.5  & 	0  & 	1.3  & 	\textbf{97.1}  & 	1.6  \\[0.2cm]
	 & \multicolumn{5}{c}{LRSM method } & 	  \multicolumn{5}{c}{PGMDL  method  } \\ 									
\cmidrule(lr){1-1}\cmidrule(lr){2-6}\cmidrule(lr){7-11}												
A & 	100  & 	\textbf{100}  & 	0  & 	0  & 	0  & 		95.3  & 	\textbf{95.3}  & 	3.9  & 	0.8  & 	0  \\
B & 	99.4  & 	0  & 	0  & 	\textbf{99.9}  & 	0.1  & 		95.1  & 	0  & 	0  & 	\textbf{95.5}  & 	4.5  \\
C & 	96  & 	0.4  & 	0  & 	\textbf{96.4}  & 	3.2  & 		95.4  & 	0  & 	0  & 	\textbf{95.5}  & 	4.5  \\
D & 	99.8  & 	0  & 	\textbf{99.8}  & 	0.2  & 	0  & 		95.2  & 	0  & 	\textbf{95.2}  & 	3.8  & 	1  \\
E & 	55.4  & 	32.3  & 	3.4  & 	\textbf{62.8}  & 	1.5  & 		42.5  & 	0.1  & 	0.6  & 	\textbf{48.2}  & 	51.1  \\
F & 	91.6  & 	0  & 	2.5  & 	\textbf{96.2}  & 	1.3  & 		70.3  & 	0  & 	0.3  & 	\textbf{73.7}  & 	26  \\[0.2cm]
	 \midrule 										
	 & \multicolumn{5}{c}{GMDL method } & 	  \multicolumn{5}{c}{SEL method} \\ 									
\cmidrule(lr){1-1}\cmidrule(lr){2-6}\cmidrule(lr){7-11}												
A & 	100  & 	\textbf{100}  & 	0  & 	0  & 	0  & 		100  & 	\textbf{100}  & 	0  & 	0  & 	0  \\
B & 	99.7  & 	0  & 	0  & 	\textbf{100}  & 	0  & 		99.7  & 	0  & 	0  & 	\textbf{100}  & 	0  \\
C & 	100  & 	0  & 	0  & 	\textbf{100}  & 	0  & 		99.8  & 	0  & 	0  & 	\textbf{100}  & 	0  \\
D & 	99.9  & 	0  & 	\textbf{99.9}  & 	0.1  & 	0  & 		99.9  & 	0  & 	\textbf{99.9}  & 	0.1  & 	0  \\
E & 	81.7  & 	2.5  & 	1.1  & 	\textbf{89.8}  & 	6.6  & 		82.7  & 	2.9  & 	1.8  & 	\textbf{92.1}  & 	3.2  \\
F & 	90.1  & 	0  & 	0.7  & 	\textbf{94.4}  & 	4.9  & 		92.9  & 	0  & 	1.1  & 	\textbf{96.9}  & 	2  \\
\bottomrule
\end{tabular}}
\label{tab:simulation}
\end{table}

Table~\ref{tab:simulation:postprob} shows the average posterior probabilities of each CP method for 
1,000 replications of the DGPs. Considering the local CP methods first (left panel), we notice that the percentage of candidate CP methods with at least two posterior probabilities above the threshold of 10\% ---mixture metric ``\# mix''--- is higher than 90\% for each DGP. This indicates that no single local method strongly dominates the others in terms of posterior probabilities. Each local method captures relevant features of the simulated data. For instance, focusing on DGP~A, BSMDL, WBDMDL, LRSM, and PGMDL are almost all equally valid methods in terms of posterior probabilities, and BS and WBS also display relatively large values for that metric. While model averaging over the potential number of breaks has already been investigated \citep[see,~\eg,][]{maheu2008learning}, we will advocate for averaging across local methods to reduce the computational cost, improve CP detection, and account for model uncertainty. Recall from Table~\ref{tab:simulation} that the SEL method appears to be an estimation strategy as accurate as Bay and Perron's global method. Table~\ref{tab:simulation:large} will further show that the computational gains with respect to the global method can be significant for long time series.

\begin{table}[H]
\centering
\spacingset{1} % DON'T change the spacing!
\caption{\textbf{Results of the simulation study -- Model posterior probabilities}\\
This table reports two metrics related to the model posterior probabilities of six local methods and the global one for the 1,000 replications of the DGPs from Table~\ref{tab:dgp}. The ``BS'', ..., ``PGMDL'', and ``$m=\cdot$'' columns show the average posterior probabilities for each DGP.  Bold values for the GMDL method highlight the true number of breaks in the corresponding DGP. ``\# mix'' designates the percentage of candidate CP methods with at least two posterior probabilities above the threshold of 10\% for each DGP.}
\scalebox{0.75}{
\begin{tabular}{c ccccc ccccccc}
\toprule 
& \multicolumn{7}{c}{Local methods} & \multicolumn{5}{c}{GMDL method}\\ 
\cmidrule(lr){2-8}\cmidrule(lr){9-13}													
DGP &  \# mix & BS   & WBS  & BSMDL  & WBSMDL & LRSM & PGMDL & \# mix & $m\!=\!0$ & $m\!=\!1$ & $m\!=\!2$ & $m\!\geq\!3$\\ 
\midrule 													
A & 100  &	8.2  & 	14.4  & 	19.6  & 	19.6  & 	19.6  & 	18.5  &0.2  & 	\textbf{99.8}  & 	0.2  & 	0.0  & 	0.0     \\
B & 98.7 &	0.2  & 	0.3  & 	16.1  & 	7.7  & 	38.2  & 	37.5          &1.0  & 	0.0  & 	0.0  & 	\textbf{99.6}  & 	0.4     \\
C & 96.1 &	2.1  & 	2.4  & 	17.9  & 	7.8  & 	27.7  & 	42.1          &1.7  & 	0.0  & 	0.0  & 	\textbf{99.2}  & 	0.8     \\
D & 98.6 &	0.7  & 	0.3  & 	0.0  & 	13.3  & 	44.6  & 	41.1          &1.1  & 	0.0  & 	\textbf{99.5}  & 	0.5  & 	0.0     \\
E & 93.5 &	3.8  & 	6.1  & 	14.6  & 	23.8  & 	30.1  & 	21.6      &29.2  & 	2.8  & 	2.0  & 	\textbf{84.1}  & 	11.1\\
F & 94.8 &	1.6  & 	1.8  & 	8.7  & 	13.4  & 	43.9  & 	30.6          &13.3  & 	0.0  & 	1.0  & 	\textbf{91.6}  & 	7.5 \\
\bottomrule
\end{tabular}}
\label{tab:simulation:postprob}
\end{table}

Regarding the global method, the right panel of Table~\ref{tab:simulation:postprob} reports the average posterior probabilities for a set of possible number of breaks. As expected, the highest average posterior probability corresponds to GMDL running with the true number of breaks (in bold). The average posterior probabilities are much more favorable to the true model when multiple values for $m$ are explored with a single global method than when multiple local methods are used. DGPs E and F display somewhat more diffuse posterior probabilities.

To underscore the usefulness of combining local CP methods in long time series, we simulate 100 series with DGP~B, varying in size from $T=2^{10} =1,\!024$ to $T=2^{14}=16,\!384$ by assuming breaks at dates $\tau_1=0.5 T$ and $\tau_2 = 0.75 T$. Table~\ref{tab:simulation:large} summarizes the simulation results. Based on the local methods' computation time, BS and WBS are the fastest algorithms, followed by BSMDL, LRSM, WBSMDL, and PGMDL. It takes about twelve minutes on a standard computer for the slowest local algorithm to run for the largest sample size.\footnote{We use an Intel Xeon Gold 6252 CPU at 2.1 GHz.} Turning to the global method, GMDL's algorithm takes more than eleven minutes to run for a sample of size $2^{12}=4,\!096$. For the largest sample sizes considered, GMDL is feasible but too demanding for the 100 replications required for our simulation study. Regarding the detection rates of the true number of CPs (Column ``$m\!=\!m_0$''), all approaches except BS and WBS deliver excellent results, as expected, given the consistency of the MDL criterion. The adapted methods proposed in Section~\ref{sec:newmethods} perform very well: The true breaks are detected in at least 95\% of the runs. BS and WBS have the fastest computation time but also the poorest performance.

When assessing whether the estimation of the breaks is accurate in Columns ``$|\widehat{\tau}_i - \tau_i|$'', BSMDL, WBSMDL, and PGMDL are as reliable as the global method. Indeed, these local methods deliver the largest average MDL values, very close to the averaged exact MDL scores computed with the GMDL method (when feasible) and larger MDL scores as compared to the LRSM method (also based on the MDL criterion). These results naturally extend to the SEL method. Overall, for this particular DGP, BSMDL is the best method as it produces fast and very accurate results. Of course, BSMDL did not perform as well for DGP~E in Table~\ref{tab:simulation}. That is why selecting the local method with the highest posterior probability is preferable in practice, as the true DGP is unknown.

\begin{table}[H]
\centering
\spacingset{1} % DON'T change the spacing!
\caption{\textbf{Results of the simulation study -- Long time series}\\
This table reports results of simulations performed with DGP~B for a long time series  by setting the relative locations at periods $\tau_1=0.5 T$ and $\tau_2 = 0.75 T$. The results are based on 100 replications. ``Time'' is the average estimation time (in minutes). ``$m\!=\!m_0$'' is the percentage of replications in which the true number of regimes has been detected. ``$|\widehat{\tau}_i - \tau_i|$'' $(i=1,2)$ is the average of the absolute value of the difference between the estimated CPs and the true one. ``$\text{MDL}$'' is the average estimated MDL value. The computational burden prevented the computation of the GMDL method on series with more than $T=2^{12}=4,\!096$ observations.}	
\scalebox{0.69}{
\begin{tabular}{c ccccc ccccc}
\toprule
$T$  & Time & $m\!=\!m_0$ & $|\widehat{\tau}_1-\tau_1|$ 	& $|\widehat{\tau}_2-\tau_2|$ &  $\text{MDL}$	 
& Time & $m\!=\!m_0$ &  $|\widehat{\tau}_1-\tau_1|$ 	& $|\widehat{\tau}_2-\tau_2|$ &  $\text{MDL}$ \\						
\midrule										
	 & \multicolumn{5}{c}{BS  method } & 	  \multicolumn{5}{c}{WBS method } \\ 								
\cmidrule(lr){1-1}\cmidrule(lr){2-6}\cmidrule(lr){7-11}											
 $2^{10}$  & 	0.00  & 	57  & 	11.37  & 	39.40  & 	-1534.43  & 	0.00  & 	63  & 	11.48  & 	70.38  & 	-1584.66  \\ 
 $2^{11}$  & 	0.00  & 	34  & 	10.00  & 	51.29  & 	-2982.35  & 	0.00  & 	57  & 	10.23  & 	72.21  & 	-2978.50  \\ 
 $2^{12}$  & 	0.01  & 	17  & 	9.82  & 	43.76  & 	-6014.65  & 	0.01  & 	60  & 	11.67  & 	59.07  & 	-6013.11  \\ 
 $2^{13}$  & 	0.01  & 	19  & 	14.37  & 	48.42  & 	-11682.57  & 	0.01  & 	57  & 	9.26  & 	44.65  & 	-11664.00  \\ 
 $2^{14}$  & 	0.03  & 	11  & 	19.73  & 	26.00  & 	-23395.69  & 	0.03  & 	54  & 	13.22  & 	40.91  & 	-23321.18  \\ 
	 & \multicolumn{5}{c}{BSMDL method } & 	  \multicolumn{5}{c}{WBSMDL method } \\ 								
\cmidrule(lr){1-1}\cmidrule(lr){2-6}\cmidrule(lr){7-11}											
 $2^{10}$  & 	0.00  & 	99  & 	6.52  & 	3.33  & 	-1500.85  & 	0.19  & 	95  & 	6.64  & 	3.35  & 	-1500.85  \\ 
 $2^{11}$  & 	0.01  & 	100  & 	4.63  & 	2.93  & 	-2967.02  & 	0.49  & 	96  & 	4.50  & 	2.89  & 	-2967.02  \\ 
 $2^{12}$  & 	0.02  & 	100  & 	5.57  & 	3.33  & 	-5979.61  & 	1.31  & 	96  & 	5.44  & 	3.41  & 	-5979.61  \\ 
 $2^{13}$  & 	0.07  & 	100  & 	5.64  & 	3.45  & 	-11636.49  & 	3.99  & 	98  & 	5.78  & 	3.51  & 	-11633.85  \\ 
 $2^{14}$  & 	0.22  & 	100  & 	4.74  & 	4.14  & 	-23308.41  & 	12.17  & 	99  & 	4.74  & 	4.15  & 	-23308.41  \\ 
	 & \multicolumn{5}{c}{LRSM method } & 	  \multicolumn{5}{c}{PGMDL  method  } \\ 								
\cmidrule(lr){1-1}\cmidrule(lr){2-6}\cmidrule(lr){7-11}											
 $2^{10}$  & 	0.01  & 	100  & 	8.26  & 	5.41  & 	-1502.28  & 	0.04  & 	99  & 	5.68  & 	5.20  & 	-1502.94  \\ 
 $2^{11}$  & 	0.05  & 	100  & 	9.43  & 	7.79  & 	-2967.51  & 	0.12  & 	100  & 	4.62  & 	3.74  & 	-2968.27  \\ 
 $2^{12}$  & 	0.17  & 	100  & 	9.33  & 	7.01  & 	-5986.78  & 	0.49  & 	100  & 	4.92  & 	3.81  & 	-5985.29  \\ 
 $2^{13}$  & 	0.64  & 	100  & 	10.08  & 	8.17  & 	-11642.16  & 	2.22  & 	100  & 	5.00  & 	4.46  & 	-11634.74  \\ 
 $2^{14}$  & 	2.45  & 	100  & 	10.42  & 	10.89  & 	-23310.94  & 	11.58  & 	100  & 	4.70  & 	5.31  & 	-23309.88  \\ 
 \midrule 									
& \multicolumn{5}{c}{GMDL method } & 	  \multicolumn{5}{c}{SEL method} \\ 								
\cmidrule(lr){1-1}\cmidrule(lr){2-6}\cmidrule(lr){7-11}											
 $2^{10}$  & 	0.45  & 	99  & 	5.34  & 	3.66  & 	-1500.85  & 	  & 	99  & 	5.97  & 	3.36  & 	-1500.85  \\ 
 $2^{11}$  & 	2.12  & 	100  & 	3.96  & 	2.90  & 	-2966.46  & 	  & 	100  & 	4.39  & 	3.00  & 	-2967.02  \\ 
 $2^{12}$  & 	11.85  & 	100  & 	4.44  & 	3.32  & 	-5978.67  & 	  & 	100  & 	5.06  & 	3.27  & 	-5979.61  \\ 
 $2^{13}$  & 	\multicolumn{5}{c}{Computationally too demanding} & 	  & 	100  & 	5.21  & 	3.39  & 	-11633.85  \\ 
 $2^{14}$  & 	\multicolumn{5}{c}{Computationally too demanding} & 	  & 	100  & 	4.16  & 	4.14  & 	-23308.41  \\ 
\bottomrule
\end{tabular}}
\label{tab:simulation:large}
\end{table}

\section{Bayesian CP Estimation and Forecasting}
\label{sec:mcmc}

To quantify the uncertainty of the estimated CPs, we propose a full Bayesian framework that accounts for information regarding the number and location of the CPs provided by an existing CP method. Our approach differs in three ways from the Bayesian frameworks of \citet{rigaill2012exact} and \citet{MaheuSong2013}, which also address break uncertainty and use $\NIG$ priors. First, we condition on a given number of breaks while they do not. Second, we can handle a broader class of priors for the CPs, such as the geometric regime duration \citep[as in][]{Chib98} or more complex distributions such as Poisson or negative binomial \citep[as in][]{KoopPotter07,bauwens2011estimating}. Third, we can use the estimated CPs to set an informative prior on the breakpoints. \new{To illustrate the latter two features, we focus on a binomial distribution that is more flexible than a Poisson distribution due to its two hyperparameters and more informative than a negative binomial distribution since its variance is smaller than its expectation.}

\new{Given a vector $\widehat{\btau}$ obtained from a CP method, we assume the following prior distribution:
\begin{equation}
\widetilde{\tau}_i \sim  \mathcal{TB}( r_i, e_i ) \,,
\end{equation}
for $i=1,\ldots,m$, where $\mathcal{TB}(r_i,e_i)$ denotes a binomial distribution with parameters $r_{i}$ and $e_{i}$, truncated on support $\left(\left\lceil \frac{\widehat{\tau}_{i-1}+\widehat{\tau}_{i}}{2}\right\rceil ,\left\lfloor  \frac{\widehat{\tau}_{i}+\widehat{\tau}_{i+1}}{2}\right\rfloor \right)$, where $\left\lceil \cdot \right\rceil$ and $\left\lfloor \cdot \right\rfloor$ are the ceiling and floor functions, respectively. The two parameters of the binomial distribution are calibrated such that $r_i=\left\lfloor  \frac{\widehat{\tau}_{i}+\widehat{\tau}_{i+1}}{2}\right\rfloor$ and $E[\widetilde{\tau}_i] = \widehat{\tau}_i$. Defining $\widetilde{\btau}=\left(\widetilde{\tau}_1,\ldots,\widetilde{\tau}_{m}\right)'$, the Bayesian model is completed as follows:
\begin{align*}
\begin{split}
y_t|\widetilde{\btau},\bbeta_i,\sigma_i^2  
& \sim   \NORM(\bx_t'\bbeta_i,\sigma_i^2)\,, \quad \text{for}\quad \widetilde{\tau}_{i-1} < t \leq \widetilde{\tau}_{i} \,, \\
\bbeta_i|\widetilde{\btau},\sigma_i^2 
& \sim  \NORM\left( \widehat{\bbeta}_i,\sigma_i^2 \ubar{g}_i (\bX_i'\bX_i)^{-1} \right)\,, \\
\sigma_i^2|\widetilde{\btau} 
& \sim \IG \left( \frac{\ubar{\nu}_i}{2},\frac{\ubar{s}_i}{2} \right) \,,
\end{split}
\end{align*}
for $i=1,\ldots,m+1$, where $\widehat{\bbeta}_i$ is the OLS estimates over segment $]\widetilde{\tau}_{i-1},\widetilde{\tau}_{i}]$ with $\widetilde{\tau}_0 = 0$, $\widetilde{\tau}_{m+1} = T$, $\ubar{\nu}_i = \sqrt{n_i}$, $\ubar{s}_i = \frac{s_i}{\sqrt{n_i}}$, and $\ubar{g}_i = \left( \frac{((m^{+})^{\frac{1}{m+1}})n_i^{\frac{1}{4}}T}{(\frac{1}{\sqrt{n_i}} +1)^{\frac{1}{2}}} \right)^{\frac{2}{K}} \!\!\!\! \exp\left( \frac{2}{K}\Delta R_{4,i} \right) n_i  - 1$ with $n_i = \widetilde{\tau}_i-\widetilde{\tau}_{i-1}$.}

\new{As shown in Proposition~\ref{prop:mlmdl}, $\bbeta_i$ and $\sigma^2_i$ can be integrated out. We can thus directly sample from the posterior distribution of the CPs conditioning on the data. Doing so significantly improves the efficiency of the sampler, as it rules out dependencies between the CPs and other model parameters. The posterior distribution of $\widetilde{\btau}$ is given by:
\begin{align}
\begin{split}\label{eq:BayesPostTau}
f( \widetilde{\btau}|\by_{1:T}) 
& \propto f^{\text{MDL}}(\by_{1:T} | \widetilde{\btau}) \prod_{i=1}^{m} 
f_{\mathcal{B}}\left(\widetilde{\tau}_i|r_i,e_i\right) 
\1{\left\lceil \frac{\widehat{\tau}_{i-1}+\widehat{\tau}_{i}}{2}\right\rceil \leq  \widetilde{\tau}_i \leq \left\lfloor  \frac{\widehat{\tau}_{i}+\widehat{\tau}_{i+1}}{2} \right\rfloor} \,,
\end{split}
\end{align}
where $f_{\mathcal{B}}\left(x|r_i,e_i\right)$ denotes the density function of the binomial distribution with parameters $r_i$ and $e_i$, and 
$\1{\bullet}$ is the indicator function. To simulate from this posterior, we use the D-DREAM algorithm \citep[][]{bauwens2011estimating}. Once the CPs have been sampled, direct sampling from the posterior distributions of $\bbeta_i$ and $\sigma^2_i$ is straightforward (see the appendix, Section~\ref{app:mcmc}).}

\new{One advantage of the Bayesian paradigm is the flexibility of the prior structure. In a CP framework, hierarchical priors have been introduced in \citet{PPT06} to learn the in-sample break process and generate informed future breaks and parameters. We show below how to adapt our framework to account for one future break.\footnote{The framework is easily extended to any number of future breaks, but it makes notations cumbersome.} We use the following prior distributions:
\begin{align}
\begin{split}\label{eq:prior_oos}
\widetilde{\tau}_{m+1}|\widetilde{\btau}  
& \sim  \Lambda \equiv T + \mathcal{G} (r_{m+1} ) \,, \\
\bbeta_{m+2}|\widehat{\bbeta}_{1:m+1} 
& \sim  \NORM\left(\frac{1}{m+1}\sum_{i=1}^{m+1}\widehat{\bbeta}_{i}, \Sigma_{\bbeta_{m+2}} \right)\,, \\
\sigma_{m+2}^2|\sigma_{1:m+1}^2 
& \sim \IG \left( \frac{T}{2},\frac{\frac{1}{m+1}\sum_{i=1}^{m+1}s_i}{2} \right) \,,
\end{split}
\end{align}
where $\mathcal{G} (r_{m+1})$ denotes a geometric distribution with break probability $r_{m+1} = \frac{m}{T}$ and $\Sigma_{\bbeta_{m+2}}$ stands for a diagonal matrix with the empirical variance of the OLS estimates as diagonal elements.\footnote{We assume implicitly that the model has more than two breakpoints since we cannot learn about the break process with less information.} Note that the deterministic transformation $\Lambda$ implies that the breakpoint occurs beyond the sample size. The out-of-sample prior specification in \eqref{eq:prior_oos} is similar to the one proposed in \citet{PPT06}. Since the information set includes all available data, this prior can be easily adapted to account for specific features such as negative correlations between parameters over consecutive segments \citep[see,~\eg,][]{pastor2001equity}  or persistence found in the OLS estimates.}

\section{Empirical Applications}
\label{sec:empirics}

\new{In this section, we propose two applications to show the practical relevance of our CP framework. The first application illustrates the computational advantage of using local CP methods when dealing with a very long time series. It stresses the ability of the MDL marginal likelihood to estimate and combine multiple CP models, accounting for break and parameter uncertainty. The second application is a forecasting exercise on the U.S. Consumer Price Index (CPI). It illustrates how combining local or global methods and incorporating an out-of-sample break can improve forecasting.}

\subsection{Combination of Local CP Methods in Fama-French Factor Models}
\label{sec:ff5}

\new{We revisit the application in \citet{HorvthEtAl2020} that seeks to detect structural changes in the exposure of a portfolio of US banking sector stocks to five Fama-French risk factors. Our application considers a longer time horizon, namely daily returns over a period ranging from July 1st, 1963 to December 30th, 2022. The sample size of 14,979 observations is substantial. The five-factor model \citep{famafrench2015} takes the following form :
\begin{equation} \label{eq:ff5}
\begin{split}
R_t - R_{\text{F},t} & = a 
+ b_{\text{MKT}} (R_{\text{MKT},t} - R_{\text{F},t} ) 
+ b_{\text{SMB}} R_{\text{SMB},t} \\
&     \qquad+ b_{\text{HML}} R_{\text{HML},t}  
 + b_{\text{RWM}} R_{\text{RWM},t} 
+ b_{\text{CMA}} R_{\text{CMA},t} 
+ \varepsilon_t \,, \quad \varepsilon_t \sim \mathcal{N}(0,\sigma^2) \,,
\end{split}
\end{equation}
where $R_t$ is the return of the bank portfolio at time $t$, $R_{\text{F},t}$ is the risk-free rate, $R_{\text{MKT},t}$ is the market return, $R_{\text{SMB},t}$ is the return on a diversified portfolio of small stocks minus the return on a diversified portfolio of big stocks; $R_{\text{HML},t}$ is the return of a portfolio of stocks with a high book-to-market (B/M) ratio minus the return of a portfolio of stocks with a low B/M ratio; $R_{\text{RMW},t}$ is the returns of a portfolio of stocks with robust profitability minus a portfolio of stocks with weak profitability; and finally $R_{\text{CMA},t}$ is the return of a portfolio of stocks with conservative investment minus the return of a portfolio of stocks with aggressive investment.}

\new{Five-factor model~\eqref{eq:ff5} is expected to exhibit evolving risk exposures over time, and some factors may not be relevant throughout the entire period. Therefore, to detect the most likely factors that impact the bank portfolio, we consider the $2^5 = 32$ possible combinations of factors and apply our local CP methods to each model. This methodology is similar to that used in \citet{meligkotsidou2008detecting} and \citet{dufays2022selective}, which also exploit different sets of explanatory variables in a CP framework to infer hedge fund investment strategies.} 

\new{For each of the 32 models, we apply eleven local CP methods. Specifically, we use the BSMDL and WBSMDL methods (both with five thresholds given by 0, 1, \ldots, 4) and the PGMDL method. The estimation time of the local CP methods is at most three minutes (for the PGMDL) on a standard computer. Once the $32 \times 11 = 352$ models are estimated, we compare their MDL marginal likelihood and select the most probable model(s). The results indicate that, in the context of this large dataset, the bank portfolio is exposed to all five risk factors since the posterior probability of Fama-French model~\eqref{eq:ff5} with all the five factors reaches 100\%.}

\begin{figure}[H]
\centering
\spacingset{1} % DON'T change the spacing!
\centering
\includegraphics[scale=.50, angle=0]{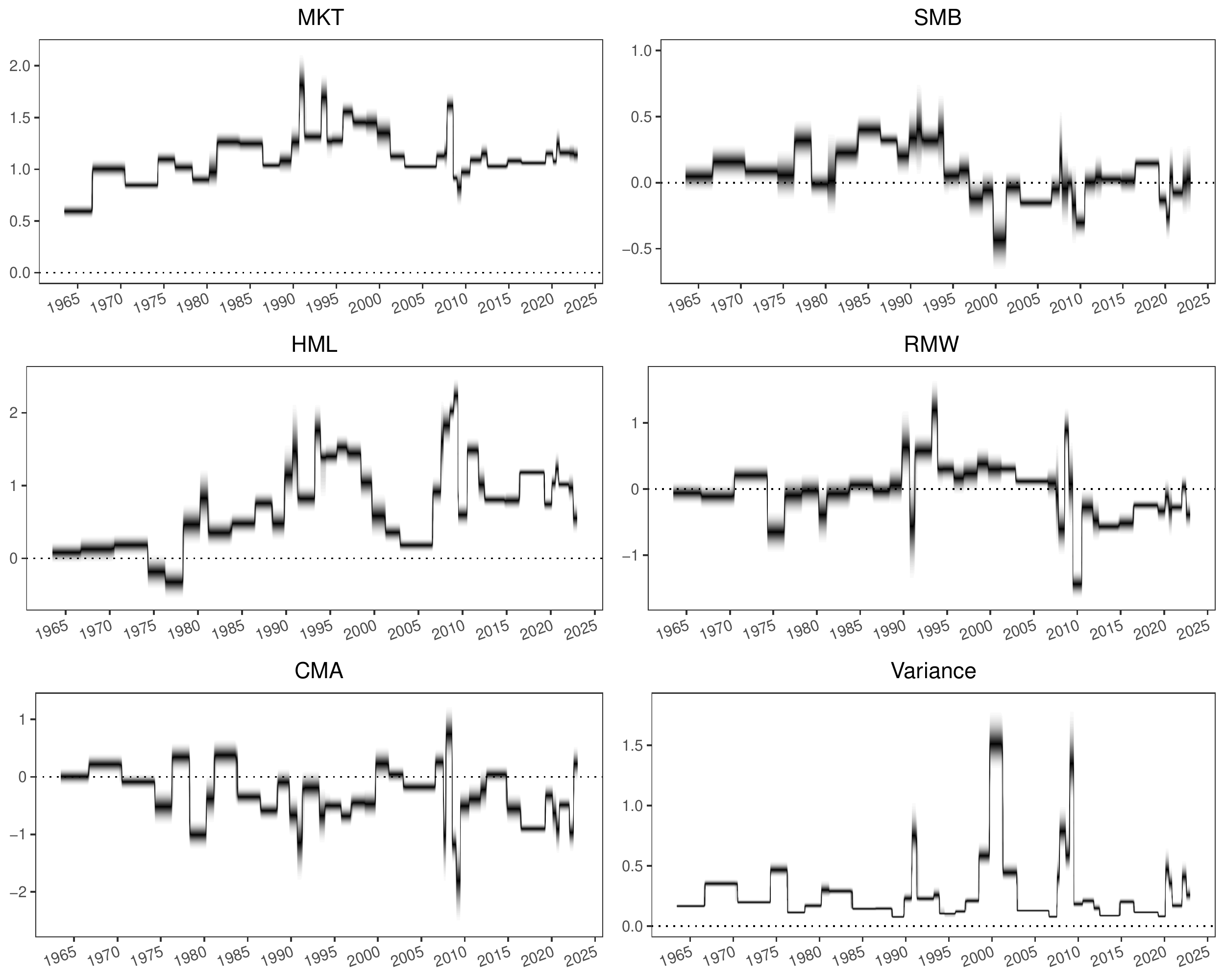} 
\caption{\textbf{Mixture of marginal posterior distributions}\\
The plots show the mixture of marginal posteriors of the Fama-French five-factor model parameters. These posteriors are computed with the full Bayesian setup of Section~\ref{sec:mcmc}. The mixing weights are the posterior probabilities of the nine CP local methods.} 
\label{fig:ff5}
\end{figure}

\new{Focusing on the model with the five factors, we apply the Bayesian setup of Section~\ref{sec:mcmc} to each of the eleven CP methods, yielding a mixture posterior that accounts for both break and method uncertainty in the model parameters. Figure~\ref{fig:ff5} displays the posterior distribution for the five-factor exposures and the variance parameters over the entire time period. We notice many breaks in the parameters. In particular, the method with the largest posterior probability (91.1\%) identifies 38 CPs, while the second-best method captures 39 breakpoints. The banking sector remained positively exposed to the market factor over the whole sample, with an exposure ranging from 0.5 to 2 and a clear increasing trend until the mid 90s. The exposure to the other  factors was sometimes positive or negative. Notice that the 2007-2008 financial crisis that sparked the Great Recession impacted all risk factors quite heavily while the COVID-19 pandemic occurring at the end of the sample seems also to be detected.}

\new{Given the significant number of CPs identified, we can gain insights into the break process and make inference about model parameters accounting for a possible future break. In particular, we examine the 38 CPs detected by the most probable model and use prior~\eqref{eq:prior_oos} to infer the bank's market exposure following a future break. Our analysis yields a 95\% credible interval of $[0.69;1.63]$ for $b_{\text{MKT}}$, indicating that the banking sector will likely remain highly exposed to the market in the upcoming period even in the case of a future breakpoint}.

\begin{figure}[H]
\centering
\spacingset{1} % DON'T change the spacing!
\centering
\includegraphics[scale=.55, angle=0]{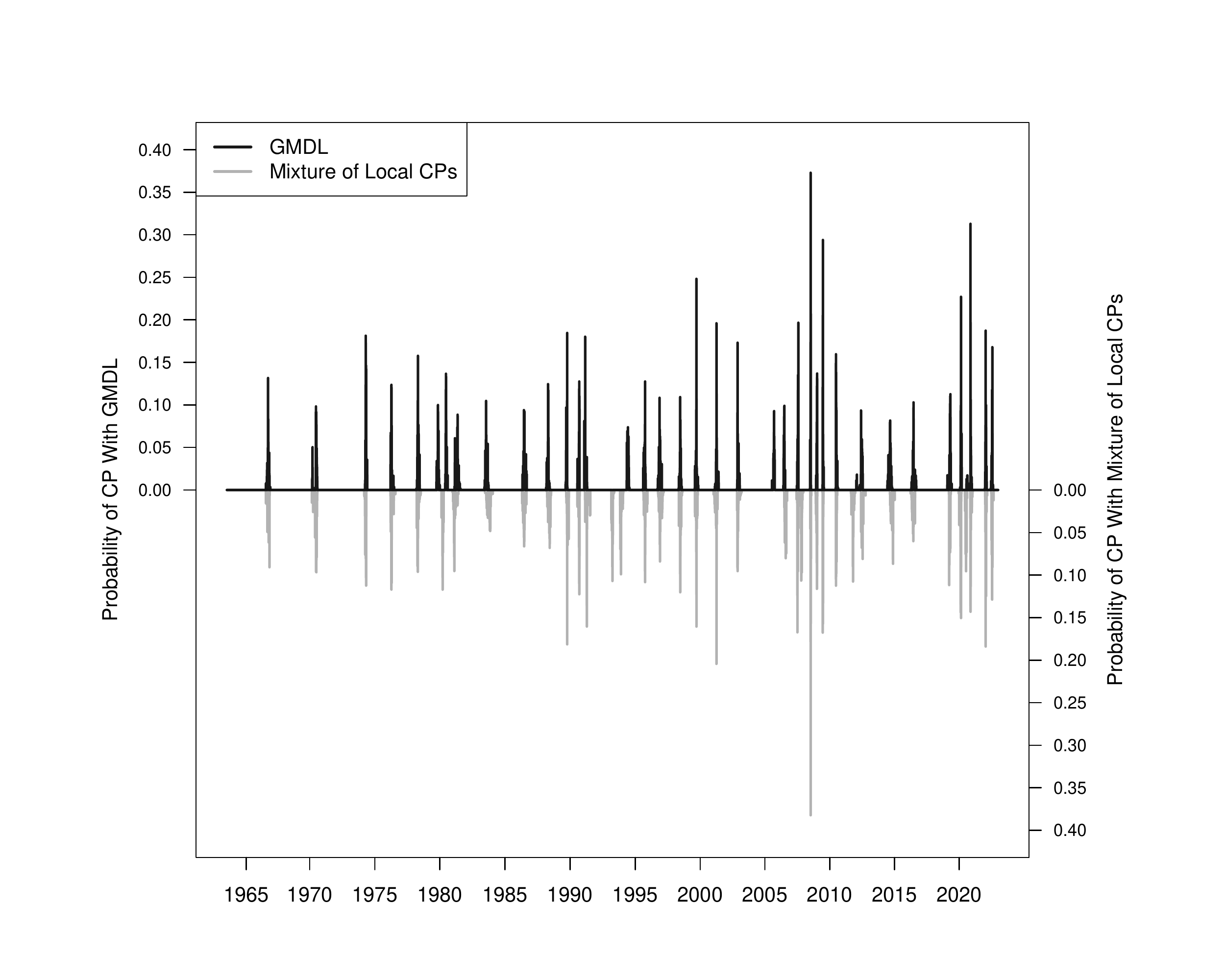} 
\caption{\textbf{Change-point probability over time}\\
The figure displays the CP probability over time obtained with the GMDL (in black) and the mixture of local CP methods (in gray).} 
\label{fig:ff5cp}
\end{figure}

\new{To verify the validity of our local CP results, we also estimate the five-factor model with the global GMDL method with a maximum of 80 regimes. The estimation procedure takes approximately two hours and identifies a number of regimes ranging from 34 to 41, with a mode at 37. We focus on the best GMDL model and examine the 36 break locations it finds. One out of 36 locations is at a distance of 230 days, five locations are within a distance of 32 days, while all the other CPs are within a range of nine days or less to the CPs identified by the local methods. Figure~\ref{fig:ff5cp} highlights the very good match between the mixture of local CPs and the GMDL in terms of the posterior probability of CPs over time.}

\subsection{Forecasting the U.S. Consumer Price Index}

\new{For the second illustration, we use our CP methods to forecast the monthly U.S. CPI. We consider the series $y_t = 1200 \times \frac{\text{CPI}_t - \text{CPI}_{t-1}}{\text{CPI}_t}$, where $\text{CPI}_t$ denotes the CPI in month $t$. The series ranges from January 1947 
to August 2022 and includes a total of 908 monthly observations. We use an expanding window scheme for the model estimation, starting with 10\% of the observations. Overall, we have 817 out-of-sample observations.
For each window, we explore both AR(1) and AR(2) specifications with and without breaks and apply local and global CP detection methods. We use the same eleven local CP methods used in the previous section. In addition, we apply the global GMDL method with a maximum of ten regimes. All models are then estimated using the Bayesian framework of Section~\ref{sec:mcmc} to generate forecasts at horizons $h \in \{1,3,6,12\}$ months.\footnote{\new{To do so, for each period and each method, we simulate from the posterior predictive distribution and average over the draws to predict the next $h$ observations. Afterward, we combine the predictions from the local or global methods using the posterior model probabilities computed from the MDL marginal log-likelihood.}}}

\new{The root mean square forecast error (RMSFE) and mean absolute forecast error (MAFE) 
are reported in Table~\ref{tab:forecasting} for the various models and forecasting horizons. In gray, we highlight the specifications that
belong to the superior set of models at the 95\% confidence level for a given loss function and forecasting 
horizon \citep{HansenEtAl2011}.\footnote{\new{We use the function \texttt{mcsTest} implemented in the R package \textbf{rugarch} \citep{rugarch}.}} Overall, for both metrics, we notice improvements when we use local or global
combination methods, in particular when we incorporate a future break in the modeling and consider forecasting over longer time horizons.}

\begin{table}[H]
\centering
\spacingset{1} % DON'T change the spacing!
\caption{\textbf{Forecasting results}\\
This table reports the root mean square forecast error (RMSFE) and mean absolute forecasting 
error (MAFE) for the various model specifications and forecasting horizons. We highlight in gray the specifications belonging to the set of superior models at the 95\% confidence level for a given loss function and forecasting horizon \citep{HansenEtAl2011}. 
The forecasting exercise is conducted over 817 out-of-sample observations.  }
\scalebox{0.8}{
\begin{tabular}{lcccccccc}
\toprule	
& \multicolumn{4}{c}{RMSFE for horizon $h=$}
& \multicolumn{4}{c}{MAFE for horizon $h=$}\\ 
\cmidrule(lr){2-5}	
\cmidrule(lr){6-9}																										
Model       & 	 1  & 	 3  & 	 6  & 	 12  & 	 1  & 	  3  & 	 6  & 	 12  \\ 
\midrule 	
AR(1) & $\grbb{3.00}$ & 3.60 & 3.76 & 3.81 & $\grbb{2.16}$ & 2.55 & 2.65 & 2.68 \\ 
AR(1) - Local & $\grbb{2.96}$ & $\grbb{3.27}$ & 3.40 & 3.57 & $\grbb{2.09}$ & $\grbb{2.31}$ & 2.40 & 2.55 \\ 
AR(1) - Local - Future Break & $\grbb{2.95}$ & $\grbb{3.27}$ & 3.39 & 3.55 & $\grbb{2.08}$ & $\grbb{2.31}$ & 2.40 & 2.54 \\ 
AR(1) - Global & $\grbb{2.95}$ & $\grbb{3.24}$ & 3.39 & 3.54 & $\grbb{2.07}$ & $\grbb{2.29}$ & 2.38 & 2.51 \\ 
AR(1) - Global - Future Break & $\grbb{2.95}$ & $\grbb{3.24}$ & $\grbb{3.38}$ & $\grbb{3.52}$ & $\grbb{2.07}$ & $\grbb{2.29}$ & $\grbb{2.37}$ & $\grbb{2.49}$ \\ 
\multicolumn{9}{c}{}\\[-.2cm]  
AR(2) & $\grbb{2.97}$ & 3.45 & 3.64 & 3.79 & $\grbb{2.12}$ & 2.45 & 2.56 & 2.67 \\ 
AR(2) - Local  & $\grbb{2.91}$ & $\grbb{3.28}$ & 3.40 & 3.56 & $\grbb{2.05}$ & $\grbb{2.32}$ & 2.40 & 2.55 \\ 
AR(2) - Local - Future Break  & $\grbb{2.91}$ & $\grbb{3.28}$ & 3.39 & 3.53 & $\grbb{2.06}$ & $\grbb{2.31}$ & 2.39 & 2.52 \\ 
AR(2) - Global & $\grbb{2.94}$ & $\grbb{3.28}$ & 3.39 & 3.55 & $\grbb{2.07}$ & $\grbb{2.31}$ & 2.38 & 2.53 \\ 
AR(2) - Global - Future Break & $\grbb{2.94}$ & $\grbb{3.28}$ & $\grbb{3.37}$ & $\grbb{3.51}$ & $\grbb{2.07}$ & $\grbb{2.30}$ & $\grbb{2.37}$ & $\grbb{2.50}$ \\
%\midrule 	
%MSE.AR1 & $\grbb{3.002}$ & 3.599 & 3.759 & 3.807 & $\grbb{2.156}$ & 2.553 & 2.655 & 2.683 \\ 
%  MSE.Local.AR1 & $\grbb{2.956}$ & $\grbb{3.275}$ & 3.398 & 3.566 & $\grbb{2.085}$ & $\grbb{2.314}$ & 2.402 & 2.551 \\ 
%  MSE.Local.AR1.FUT & $\grbb{2.953}$ & $\grbb{3.274}$ & 3.394 & 3.546 & $\grbb{2.082}$ & $\grbb{2.313}$ & 2.396 & 2.536 \\ 
%  MSE.Global.AR1 & $\grbb{2.948}$ & $\grbb{3.239}$ & 3.389 & 3.543 & $\grbb{2.071}$ & $\grbb{2.291}$ & 2.383 & 2.509 \\ 
%  MSE.Global.AR1.FUT & $\grbb{2.948}$ & $\grbb{3.237}$ & $\grbb{3.380}$ & $\grbb{3.517}$ & $\grbb{2.069}$ & $\grbb{2.287}$ & $\grbb{2.376}$ & $\grbb{2.487}$ \\ 
%  MSE.AR2 & $\grbb{2.975}$ & 3.449 & 3.638 & 3.787 & $\grbb{2.120}$ & 2.445 & 2.558 & 2.668 \\ 
%  MSE.Local.AR2 & $\grbb{2.913}$ & $\grbb{3.283}$ & 3.397 & 3.56 & $\grbb{2.054}$ & $\grbb{2.316}$ & 2.397 & 2.546 \\ 
%  MSE.Local.AR2.FUT & $\grbb{2.913}$ & $\grbb{3.278}$ & 3.389 & 3.531 & $\grbb{2.055}$ & $\grbb{2.313}$ & 2.391 & 2.521 \\ 
%  MSE.Global.AR2 & $\grbb{2.941}$ & $\grbb{3.284}$ & 3.389 & 3.553 & $\grbb{2.071}$ & $\grbb{2.306}$ & 2.381 & 2.531 \\ 
%  MSE.Global.AR2.FUT & $\grbb{2.939}$ & $\grbb{3.280}$ & $\grbb{3.373}$ & $\grbb{3.508}$ & $\grbb{2.069}$ & $\grbb{2.303}$ & $\grbb{2.369}$ & $\grbb{2.499}$ \\
\bottomrule
\end{tabular}}
\label{tab:forecasting}
\end{table}

\section{Conclusion}
\label{sec:conclusion}

Many methods are available for detecting structural breaks and estimating the corresponding model parameters in each regime in linear regression models where the number and locations of the breaks are unknown. These methods mainly differ in the statistical framework under which they are developed (\ie, Bayesian or frequentist), the fitting criterion used to adjust the data, and the algorithm's complexity used to optimize the fitting criterion. Even the most reliable global detection methods can deliver quite different numbers and locations of CPs when applied to real-world data. Moreover, these methods are computationally demanding for large samples. 

This article addresses both issues by focusing on normal-linear regression models. We propose a Bayesian framework that allows researchers to account for model and break uncertainty and obtain accurate and reliable estimates for large data sets at a low computational cost. \new{Our methodology relies on a marginal likelihood that exactly corresponds to the MDL criterion widely used in the frequentist literature for CP detection in linear regression. We establish that a class of hyperparameters for the $\NIG$ prior distribution achieves such an equivalence.} We further propose one global and three local CP methods that build on the probabilistic interpretation of the MDL criterion and on recent CP algorithms proposed in the literature. We also explore a full Bayesian model that delivers credible intervals for the CPs and other quantities of interest. Extensive experiments based on simulated data demonstrate the excellent performance of the proposed approaches to detect CPs relative to other state-of-the-art methods. \new{Finally, we provide empirical illustrations that show the practical relevance of our approach when dealing with long time series and forecasting.}

Future research could consider developing this framework in multivariate piecewise-linear regressions. For instance, \citet{Smith2022} uses a Normal Inverse-Wishart prior distribution to obtain a closed-form expression of the marginal likelihood given the breakpoints. We could extend his framework by calibrating those prior hyperparameters to match a consistent information criterion and using the efficient breakpoint sampler proposed in this paper.

% REFERENCES
\newpage
\spacingset{1} % DON'T change the spacing!
%\footnotesize
\bibliographystyle{elsarticle-harv-doi}

\spacingset{1} % DON'T change the spacing!
\newpage
\begin{titlepage}
\begin{center}
\vspace*{1cm}
\huge{
Linking Frequentist and Bayesian\\
Change-Point Methods\\[.5cm]
Appendix}\\   
%\vspace{1.5cm}
%\Large{
%David Ardia$^{1}$,
%Arnaud Dufays$^{2}$,
%Carlos Ord\'as Criado$^{3}$}\\[.5cm]
%\normalsize{
%$^{1}$GERAD \& Department of Decision Sciences, HEC Montr\'eal, Canada\\
%$^{2}$Faculty of Data Science, Economics \& Finance, EDHEC Business School, France\\
%$^{3}$Department of Economics, Laval University, Canada}
\vfill
\end{center}
\end{titlepage}

\newpage
\clearpage
\pagenumbering{arabic} 
\linespread{1.2}
%\onehalfspacing

\appendix 
\setlength\parindent{0pt}

\setcounter{propositionapp}{0}
\renewcommand*{\thepropositionapp}{A.\arabic{propositionapp}} %Alph alph

\setcounter{remarkapp}{0}
\renewcommand*{\theremarkapp}{A.\arabic{remarkapp}} 

\setcounter{equation}{0}
\renewcommand*{\theequation}{A\arabic{equation}} %Alph alph

\setcounter{table}{0}
\renewcommand*{\thetable}{A.\arabic{table}} 

\setcounter{section}{0}
\renewcommand*{\thesection}{\Roman{section}} 

\setcounter{subsection}{0}
\renewcommand*{\thesubsection}{\thesection.\Alph{subsection}} 

\setcounter{page}{1}
\renewcommand*{\thepage}{\arabic{page}} 

\newpage
\section{Proofs of the Propositions}
\small

\subsection{Proofs of Proposition~\ref{prop:mlnig}}
\label{app:prop:mlnig}

\new{%
The $\NIG$ priors are given by:
\begin{align} 
f(\bbeta_i | \sigma_i^2, \btau) 
& = (2\pi)^{-\frac{K}{2}}|\sigma_i^2 \ubar{g}_i \ubar{\bM}_i^{-1}|^{-\frac{1}{2}}\exp \left(-\frac{1}{2\sigma_i^2}(\bbeta_i-\ubar{\bbeta}_i)'\ubar{g}_i^{-1} \ubar{\bM}_i(\bbeta_i-\ubar{\bbeta}_i)\right)  \,, \\
f(\sigma_i^2|\btau) 
& = \frac{(\frac{\ubar{s}_i}{2})^{\frac{\ubar{\nu}_i}{2}}}{\Gamma(\frac{\ubar{\nu}_i}{2})} (\sigma_i^2)^{-\frac{\ubar{\nu}_i+2}{2}}\exp \left(-\frac{\ubar{s}_i}{2\sigma_i^2}\right) \,.
\end{align}
The normalizing constant of the NIG kernel is:
\begin{align}
\begin{split} 
C(\ubar{g}_i \ubar{\bM}_i^{-1},\frac{\ubar\nu_i}{2},\frac{\ubar s_i}{2}) & = \int \int (\sigma_i^2)^{-(\ubar{\nu}_i+2)/2}\exp \left(-\frac{\ubar{s}_i}{2\sigma_i^2}\right) \\
& ~~~~~~~~ (\sigma_i^2)^{-K/2}\exp \left(-\frac{1}{2\sigma_i^2}[(\bbeta_i-\ubar{\bbeta}_i)'\ubar{g}_i^{-1} \ubar{\bM}_i(\bbeta_i-\ubar{\bbeta}_i)]\right) d\bbeta_i d\sigma_i^2 \,,\\
& =  (2\pi)^{\frac{K}{2}}|\ubar{g}_i \ubar{\bM}_i^{-1}|^{\frac{1}{2}} \frac{\Gamma(\frac{\ubar{\nu}_i}{2})}{(\frac{\ubar{s}_i}{2})^{\frac{\ubar{\nu}_i}{2}}}\,.
\end{split} 
\end{align}
The likelihood function over the segment $i$ is given by:
\begin{equation} 
f(\by_i|\btau,\bbeta_i,\sigma_i^2) 
= (2\pi \sigma_i^2)^{-\frac{n_i}{2}}\exp \left(-\frac{1}{2\sigma_i^2}[s_i + (\bbeta_i-\widehat{\bbeta}_i)'\bX_i'\bX_i(\bbeta_i-\widehat{\bbeta}_i)]\right)\,,
\end{equation}
where $s_i  = (\by_i - \bX_i\widehat{\bbeta}_i)'(\by_i - \bX_i\widehat{\bbeta}_i)$.
Using the above expressions, the marginal likelihood reads as follows:
\begin{align}\label{eq:MLL_dev} 
\begin{split} 
f(\by_i| \btau)& = C(\ubar{g}_i \ubar{\bM}_i^{-1},\frac{\ubar\nu_i}{2},\frac{\ubar s_i}{2})^{-1} \int \int (2\pi)^{-\frac{n_i}{2}} (\sigma_i^2)^{-(n_i+\ubar{\nu}_i+K+2)/2}\\
 &  \exp \left(-\frac{1}{2\sigma_i^2}[\ubar{s}_i+s_i + \underbrace{(\bbeta_i-\ubar\bbeta_i)'\ubar{g}_i^{-1} \ubar{\bM}_i(\bbeta_i-\ubar\bbeta_i) + (\bbeta_i-\widehat{\bbeta}_i)'\bX_i'\bX_i(\bbeta_i-\widehat{\bbeta}_i)}_{=F}]\right) d\bbeta_i d\sigma_i^2  \,.
\end{split} 
\end{align}
Focusing on $F$, we can collect terms $\bbeta_i, \ubar{\bbeta}_i$, and $\widehat{\bbeta}_i$ as follows:
\begin{align} 
F & = \bbeta_i '(\underbrace{\ubar{g}_i^{-1} \ubar{\bM}_i + \bX_i'\bX_i}_{=\bar{\bM}_i})\bbeta_i -2 \bbeta_i'\big(\underbrace{\ubar{g}_i^{-1} \ubar{\bM}_i\ubar{\bbeta}_i + \bX_i'\bX_i\widehat{\bbeta}_i}_{={\bar{\bM}}_i \bar{\bbeta}_i} \big) + \ubar{\bbeta}_i'\ubar{g}_i^{-1} \ubar{\bM}_i \ubar{\bbeta}_i  + \widehat{\bbeta}_i'\bX_i'\bX_i\widehat{\bbeta}_i \,,\label{appeq:abcd}
\end{align}
where we introduce the following notation:
\begin{align} 
\bar{\bM}_i & = \ubar{g}_i^{-1} \ubar{\bM}_i + \bX_i'\bX_i \,, \label{appeq:a} \\
\bar{\bM}_i\bar{\bbeta}_i & = \ubar{g}_i^{-1} \ubar{\bM}_i \ubar{\bbeta}_i + \bX_i'\bX_i \widehat{\bbeta}_i \,,\\
\bar{\bbeta}_i & = \bar{\bM}_i^{-1}\big(\ubar{g}_i^{-1} \ubar{\bM}_i \ubar{\bbeta}_i + \bX_i'\bX_i \widehat{\bbeta}_i\big). \label{appeq:c}
\end{align}
By using \eqref{appeq:a} to \eqref{appeq:c}, we can thus rewrite $F$ as:
\begin{align}
F & = \bbeta_i'\bar{\bM}_i \bbeta_i + \ubar{\bbeta}_i'\ubar{g}_i^{-1} \ubar{\bM}_i \ubar{\bbeta}_i - 2\bbeta_i'\bar{\bM}_i\bar{\bbeta}_i + \widehat{\bbeta}_i'\bX_i'\bX_i\widehat{\bbeta}_i\, \nonumber  \\
& = \ubar{\bbeta}_i'\ubar{g}_i^{-1} \ubar{\bM}_i \ubar{\bbeta}_i + \widehat{\bbeta}_i'\bX_i'\bX_i\widehat{\bbeta}_i - \bar{\bbeta}_i'\bar{\bM}_i\bar{\bbeta}_i + \bbeta_i'\bar{\bM}_i \bbeta_i - 2\bbeta_i'\bar{\bM}_i\bar{\bbeta}_i + \bar{\bbeta}_i'\bar{\bM}_i\bar{\bbeta}_i \, \nonumber \\ 
& = \ubar{\bbeta}_i'\ubar{g}_i^{-1} \ubar{\bM}_i \ubar{\bbeta}_i + \widehat{\bbeta}_i'\bX_i'\bX_i\widehat{\bbeta}_i - \bar{\bbeta}_i'\bar{\bM}_i\bar{\bbeta}_i + (\bbeta_i - \bar{\bbeta}_i)'\bar{\bM}_i(\bbeta_i - \bar{\bbeta}_i) \,. \label{eq:F} 
\end{align}
Using \eqref{eq:F} in marginal likelihood \eqref{eq:MLL_dev} leads to:
\begin{align} 
\begin{split} 
f(\by_i | \btau)& = C(\ubar{g}_i \ubar{\bM}_i^{-1},\frac{\ubar\nu_i}{2},\frac{\ubar s_i}{2})^{-1} \int \int (2\pi)^{-\frac{n_i}{2}} (\sigma_i^2)^{-(n_i+\ubar{\nu}_i+2)/2}\\
 &  ~~ \exp \left(-\frac{1}{2\sigma_i^2}[\ubar{s}_i+s_i + \ubar{\bbeta}_i'\ubar{g}_i^{-1} \ubar{\bM}_i \ubar{\bbeta}_i + \widehat{\bbeta}_i'\bX_i'\bX_i\widehat{\bbeta}_i - \bar{\bbeta}_i'\bar{\bM}_i\bar{\bbeta}_i]\right) \\
 & ~~~~ (\sigma_i^2)^{-\frac{K}{2}}\exp \left(-\frac{1}{2\sigma_i^2}[(\bbeta_i - \bar{\bbeta}_i)'\bar{\bM}_i(\bbeta_i - \bar{\bbeta}_i)]\right) d\bbeta_i d\sigma_i^2 \\
 & = C(\ubar{g}_i \ubar{\bM}_i^{-1},\frac{\ubar\nu_i}{2},\frac{\ubar s_i}{2})^{-1} (2\pi)^{-\frac{n_i-K}{2}} |\bar{\bM}_i^{-1}|^{\frac{1}{2}} \\
 & \int (\sigma_i^{-2})^{(n_i+\ubar{\nu}_i+2)/2}\exp \left(-\frac{1}{2\sigma_i^2}[\ubar{s}_i+s_i + \ubar{\bbeta}_i'\ubar{g}_i^{-1} \ubar{\bM}_i \ubar{\bbeta}_i + \widehat{\bbeta}_i'\bX_i'\bX_i\widehat{\bbeta}_i - \bar{\bbeta}_i'\bar{\bM}_i\bar{\bbeta}_i]\right)d\sigma_i^2 \\
& = C(\ubar{g}_i \ubar{\bM}_i^{-1},\frac{\ubar\nu_i}{2},\frac{\ubar s_i}{2})^{-1} (2\pi)^{-\frac{n_i-K}{2}} |\bar{\bM}_i^{-1}|^{\frac{1}{2}} \frac{\Gamma(\frac{\bar{\nu}_i}{2})}{\frac{\bar{s}_i}{2}^{\frac{\bar{\nu}_i}{2}}} \\
& = (2\pi)^{-\frac{n_i}{2}} \left(\frac{|\bar{\bM}_i^{-1}|}{|\ubar{g}_i \ubar{\bM}_i^{-1}|} \right)^{\frac{1}{2}} \frac{\Gamma(\frac{\bar{\nu}_i}{2})}{\Gamma(\frac{\ubar{\nu}_i}{2})}    \frac{(\frac{\ubar{s}_i}{2})^{\frac{\ubar{\nu}_i}{2}}}{(\frac{\bar{s}_i}{2})^{\frac{\bar{\nu}_i}{2}}} \,,
\end{split} 
\end{align}
with $\bar{\nu}_i = n_i+\ubar{\nu}_i$ and $\bar{s}_i = \ubar{s}_i+s_i + \ubar{\bbeta}_i'\ubar{g}_i^{-1} \ubar{\bM}_i \ubar{\bbeta}_i + \widehat{\bbeta}_i'\bX_i'\bX_i\widehat{\bbeta}_i - \bar{\bbeta}_i'\bar{\bM}_i\bar{\bbeta}_i$.}

\subsection{Proofs of Proposition~\ref{prop:mlmdl}}
\label{app:prop:mlmdl}

\new{%
If $\ubar{\bM}_i = \bX_i'\bX_i$ and $\ubar{\bbeta}_i = \widehat{\bbeta}_i$, we have:
\begin{align} 
\bar{\bM}_i & = (1+\ubar{g}_i^{-1}) \bX_i'\bX_i\,, \\
\bar{\bM}_i\bar{\bbeta}_i & = (1+\ubar{g}_i^{-1}) \bX_i'\bX_i \widehat{\bbeta}_i \,, \\
\bar{\bbeta}_i & = \widehat{\bbeta}_i \,,\\
\begin{split} 
\bar{s}_i & = \ubar{s}_i+s_i + \widehat{\bbeta}_i'\ubar{g}_i^{-1} \bX_i'\bX_i \widehat{\bbeta}_i+ \widehat{\bbeta}_i'\bX_i'\bX_i\widehat{\bbeta}_i - \widehat{\bbeta}_i'(1+\ubar{g}_i^{-1}) \bX_i'\bX_i \widehat{\bbeta}_i \, \\
 & = \ubar{s}_i+s_i + \widehat{\bbeta}_i'(\ubar{g}_i^{-1} + 1 - 1 - \ubar{g}_i^{-1})\bX_i'\bX_i \widehat{\bbeta}_i \\
 & = \ubar{s}_i+s_i \,. 
\end{split} 
\end{align}
We can simplify the marginal likelihood of segment $i$ as follows:
\begin{equation} 
\begin{split} 
f(\by_i|\btau)
= (2\pi)^{-\frac{n_i}{2}} \left( \frac{1}{1+\ubar{g}_i} \right)^{\frac{K}{2}} \frac{\Gamma(\frac{\bar{\nu}_i}{2})}{\Gamma(\frac{\ubar{\nu}_i}{2})}    \frac{(\frac{\ubar{s}_i}{2})^{\frac{\ubar{\nu}_i}{2}}}{(\frac{\bar{s}_i}{2})^{\frac{\bar{\nu}_i}{2}}} \,.
\end{split} 
\end{equation}
The marginal log-likelihood of segment $i$ reads:
\begin{equation} 
\ln f(\by_i|\btau) 
= -\frac{n_i}{2} \ln(2\pi) 
- \frac{K}{2} \ln \left(1+\ubar{g}_i\right) 
+ \ln \Gamma \left(\frac{\bar{\nu}_i}{2} \right) 
- \ln \Gamma \left(\frac{\ubar{\nu}_i}{2} \right) 
+ \ln \left(\frac{\ubar{s}_i}{2}\right)^{\frac{\ubar{\nu}_i}{2}} 
+ \ln \left(\frac{\bar{s}_i}{2}\right)^{-\frac{\bar{\nu}_i}{2}}\,. 
\end{equation}
Recalling Stirling's approximation of the gamma function for real $x>0$, we have:
\begin{equation} %\label{eq:ww}
\ln \Gamma(x) = x\ln x - x -\frac{1}{2} \ln x + \frac{1}{2}\ln (2\pi) + R_N(x) + \mathcal{O}\left(x^{-(2N-1)}\right)\,,
\end{equation}
with $R_N(x)  = \sum_{n=1}^{N-1} \frac{B_{2n}}{2n(2n-1)x^{2n-1}}$ in which $B_{2n}$ denotes the Bernoulli numbers \citep[\eg,][]{Nemes2015}. Applying the Stirling's approximation as well as substituting $\ubar{\nu}_i=\ubar{k}_i n_i$ and $\ubar s_i = \ubar{k}_i s_i$ , we find that:
\begin{align} \label{eq:ww}
\scriptsize
\begin{split}
\underbrace{\ln \Gamma \left(\frac{\bar{\nu}_i}{2}\right)}_{A_1}
&= \frac{n_i(1+\ubar{k}_i)}{2}\ln \frac{n_i(1+\ubar{k}_i)}{2} 
- \frac{n_i(1+\ubar{k}_i)}{2} 
- \frac{1}{2} \ln (n_i(1+\ubar{k}_i)) 
+ \frac{1}{2}\ln (4\pi) 
+ R_N \left(\frac{n_i+\ubar{\nu}_i}{2} \right) 
+ \mathcal{O}\left((n_i+\ubar{\nu}_i)^{-(2N-1)}\right)\,, \\
\underbrace{\ln \Gamma \left( \frac{\ubar \nu_i}{2} \right)}_{A_2}
&= \frac{\ubar{k}_i n_i}{2} \ln \frac{\ubar{k}_i n_i}{2} 
- \frac{\ubar{k}_i n_i}{2} 
- \frac{1}{2} \ln (\ubar{k}_i n_i) 
+ \frac{1}{2}\ln (4\pi) 
+ R_N \left(\frac{\ubar{k}_i n_i}{2} \right) 
+ \mathcal{O}\left((\ubar{k}_i n_i)^{-(2N-1)}\right) \\
& = \frac{n_i}{2}\ln \frac{n_i}{2} 
+ \frac{n_i}{2}\ln (1 + \ubar{k}_i) 
+ \frac{n_i \ubar{k}_i}{2}\ln \frac{n_i(1 + \ubar{k}_i)}{2} - \frac{n_i}{2} - \frac{n_i \ubar{k}_i}{2} \\
& ~~~~~~~~  - \frac{1}{2} \ln n_i 
- \frac{1}{2} \ln (\ubar{k}_i +1) 
+ \frac{1}{2}\ln (4\pi) 
+ R_N \left(\frac{n_i+\ubar{\nu}_i}{2}\right) 
+ \mathcal{O}\left((n_i+\ubar{\nu}_i)^{-(2N-1)}\right)\,, \\
\underbrace{\ln \left( \frac{\ubar s_i}{2} \right)^{\frac{\ubar \nu_i}{2}}}_{A_3} 
& = \frac{\ubar{k}_i n_i}{2} \ln \frac{\ubar{k}_i s_i}{2} \,, \\
\underbrace{\ln \left( \frac{\bar{s}_i}{2} \right)^{-\frac{\bar{v}_i}{2}}}_{A_4} 
& = -\frac{n_i(1 + \ubar{k}_i)}{2} \ln \frac{s_i(1 + \ubar{k}_i)}{2}  
= - \frac{n_i}{2}\ln \frac{s_i}{2} 
- \frac{n_i}{2}\ln (1 + \ubar{k}_i)  
- \frac{n_i \ubar{k}_i}{2} \ln \frac{s_i(1 + \ubar{k}_i)}{2} \,. \\
\end{split}
\end{align}
Using the terms in \eqref{eq:ww}, neglecting the approximation order for convenience, and 
recalling that $\ln f(\by_i|\widehat\Theta_{\text{MLE}},\btau) = -\frac{n_i}{2}\ln(2\pi \frac{s_i}{n_i}) - \frac{n_i}{2}$, we get:
\begin{equation*} 
\scriptsize
A_1-A_2+A_3+A_4 
= \ln f(\by_i|\widehat\Theta_{\text{MLE}},\btau)
+ \frac{n_i}{2}\ln(2\pi) 
+ \frac{1}{2} \ln \frac{\ubar{k}_i}{\ubar{k}_i + 1} 
+ R_N \left( \frac{n_i+\ubar{\nu}_i}{2} \right) 
- R_N \left( \frac{\ubar{k}_i n_i}{2} \right) \,, 
\end{equation*}
which yields:
\begin{equation*}
\ln f(\by_i|\btau) 
= \ln f(\by_i|\widehat\Theta_{\text{MLE}},\btau) -\frac{K}{2} \ln (1+ \ubar g_i) + \frac{1}{2} \ln \frac{\ubar{k}_i}{\ubar{k}_i +1} + \Delta R_{N,i} \,, \\
\end{equation*}
where $\Delta R_{N,i} = R_N(\frac{n_i+\ubar{\nu}_i}{2}) - R_N(\frac{\ubar{k}_i n_i}{2})$. Summing over all segments and setting $\ubar{k}_i=\frac{1}{\sqrt{n_i}}$, $\ubar g_i = \ubar{f}_{\!i} n_i - 1$ with $\ubar{f}_{\!i} = \left(\frac{((m^{+})^{\frac{1}{m+1}})n_i^{\frac{1}{4}}T}{(\frac{1}{\sqrt{n_i}} +1)^{\frac{1}{2}}}\right)^{\frac{2}{K}} \!\!\!\! \exp(\frac{2}{K}\Delta R_{N,i})$, we find:
\begin{align} %\label{eq:ww}
\scriptsize
\begin{split}
\sum_{i=1}^{m+1}\ln f(\by_i|\btau)  
& = \ln f(\by_{1:T}|\widehat\Theta_{\text{MLE}},\btau) 
- \frac{K}{2} \sum_{i=1}^{m+1} \ln n_i 
- \frac{K}{2} \sum_{i=1}^{m+1} \ln \ubar{f}_{\!i} 
-  \frac{1}{2}  \sum_{i=1}^{m+1} \ln \frac{\frac{1}{\sqrt{n_i}} + 1}{\frac{1}{\sqrt{n_i}}} 
+ \sum_{i=1}^{m+1}\Delta R_{N,i} \\
& = \ln f(\by_{1:T}|\widehat\Theta_{\text{MLE}},\btau) 
- \frac{K+1}{2} \sum_{i=1}^{m+1} \ln n_i\\ 
& ~~~~~~~~  - \frac{K}{2} \sum_{i=1}^{m+1} \ln \left(\frac{((m^{+})^{\frac{1}{m+1}})T}{\frac{1}{(\sqrt{n_i}} + 1)^{\frac{1}{2}}}\right)^{\frac{2}{K}}  \!\! \exp \left (\frac{2}{K}\Delta R_{N,i} \right)
-  \frac{1}{2}  \sum_{i=1}^{m+1} \ln \left( \frac{1}{\sqrt{n_i}} +1 \right) 
+ \sum_{i=1}^{m+1}\Delta R_{N,i} \\
& = \ln f(\by_{1:T}|\widehat\Theta_{\text{MLE}},\btau) 
- \frac{K+1}{2} \sum_{i=1}^{m+1} \ln n_i 
- \ln^{\!+}\!(m) 
- \frac{K}{2} \sum_{i=1}^{m+1} \ln (T (\exp(\Delta R_{N,i})))^{\frac{2}{K}}  + \sum_{i=1}^{m+1}\ln (\exp(\Delta R_{N,i})) \\
& = \ln f(\by_{1:T}|\btau,\widehat \Theta_{\text{MLE}})  
- \ln^{\!+}\!(m) - (m+1)\ln T - \left( \frac{K+1}{2} \right) \sum_{i=1}^{m+1}\ln n_i \\
& = \text{MDL}(m,\btau) \, .
\end{split}
\end{align}
The last equality holds up to an approximation order of $\sum_{i=1}^{m+1} \mathcal{O}\left((\ubar{k}_i n_i)^{-(2N-1)}\right)$. When $N=4$, it is bounded by $\mathcal{O}\left(\min_i(n_i)^{-\frac{7}{2}}\right)$ and this precision is sufficient in most applications.}

\subsection{Proof of Remark~\ref{rem:mle}}
\label{app:rem:mle}

\new{%
For $\ubar{k}_i=\frac{1}{\sqrt{n_i}}$, we have $\ubar{v}_i = \ubar{k}_i n_i = \sqrt{n_i}$ and $\ubar{s}_i = \ubar{k}_i s_i = \frac{s_i}{\sqrt{n_i}}$. Since $\sigma_i^2|\btau \sim IG(\frac{\ubar{v}_i}{2},\frac{\ubar{s}_i}{2})$, we have:
\begin{equation} %\label{eq:ww}
\begin{split}
E[\sigma_i^2|\btau] &= \frac{\ubar{s}_i}{\ubar{v}_i-2} = \frac{\frac{s_i}{\sqrt{n_i}}}{\sqrt{n_i}-2} = \frac{s_i}{\sqrt{n_i}(\sqrt{n_i} - 2)} = \frac{s_i}{n_i - 2\sqrt{n_i}} ~~ > \sigma_{\text{MLE}}^2 = \frac{s_i}{n_i} \,,\\
\text{Mode} & = \frac{s_i}{\sqrt{n_i}(\sqrt{n_i} + 2)} = \frac{s_i}{n_i + 2\sqrt{n_i}} ~~ < \sigma_{\text{MLE}}^2 = \frac{s_i}{n_i} \,.\\
\end{split}
\end{equation}}

%\newpage
\section{Posteriors of the Bayesian Estimation in Section~\ref{sec:mcmc}}
\label{app:mcmc}

\new{%
The full conditional distribution of the Gibbs sampler is given by:
\begin{align} %\label{eq:ww}
\scriptsize
\begin{split}
f(\bbeta_i,\sigma_i^2|y_{1:T},\btau) 
& \propto f(\by_i|\btau,\bbeta_i,\sigma_i^2)f(\bbeta_i|\sigma_i^2,\btau)f(\sigma_i^2|\btau) \\
& \propto (\sigma_i^2)^{-\frac{n_i}{2}} (\sigma_i^2)^{-\frac{K}{2}}(\sigma_i^2)^{-(\frac{\ubar{\nu}_i +2}{2})}\exp\left(-\frac{1}{2\sigma_i^2}(s_i + (\bbeta_i-\hat{\bbeta}_i)'(1+\ubar{g}_i^{-1})(\bX_i'\bX_i)(\bbeta_i-\hat{\bbeta}_i))  -\frac{\ubar{s}_i}{2\sigma_i^2} \right) \\
& = (\sigma_i^2)^{\frac{-n_i-\ubar{\nu}_i -2}{2}}  (\sigma_i^2)^{-\frac{K}{2}}\exp\left(-\frac{1}{2}(\bbeta_i-\hat{\bbeta}_i)'\frac{(1+\ubar{g}_i^{-1})}{\sigma_i^2}(\bX_i'\bX_i)(\bbeta_i-\hat{\bbeta}_i)  -\frac{\ubar{s}_i + s_i}{2\sigma_i^2} \right)  \\
& = (\sigma_i^2)^{\frac{-(n_i+\ubar{\nu}_i)}{2} - 1} \exp\left(-\frac{\ubar{s}_i + s_i}{2\sigma_i^2}\right) (\sigma_i^2)^{-\frac{K}{2}}\exp\left(-\frac{1}{2}(\bbeta_i-\hat{\bbeta}_i)'\frac{(1+\ubar{g}_i^{-1})}{\sigma_i^2}(\bX_i'\bX_i)(\bbeta_i-\hat{\bbeta}_i) \right) \,.
\end{split}
\end{align}
Direct sampling is therefore achieved as follows:
\begin{align} %\label{eq:ww}
\footnotesize
\begin{split}
\sigma_i^2|y_{1:T},\btau 
& \sim  \IG\left(\frac{n_i+\ubar{\nu}_i}{2},\frac{\ubar{s}_i + s_i}{2}\right) \,, \\
\bbeta_{i}|y_{1:T},\btau,\sigma_i^2 
& \sim  \NORM\left(\widehat{\bbeta}_i, \frac{\sigma_i^2}{(1+\ubar{g}_i^{-1})}(\bX_i'\bX_i)^{-1} \right)\,,
\end{split}
\end{align}
with $\ubar{\nu}_i = \sqrt{n_i}$, $\ubar{s}_i = \frac{s_i}{\sqrt{n_i}}$ and $\ubar{g}_i = \left(\frac{((m^{+})^{\frac{1}{m+1}})n_i^{\frac{1}{4}}T}{(\frac{1}{\sqrt{n_i}} +1)^{\frac{1}{2}}}\right)^{\frac{2}{K}} \!\!\!\! \exp(\frac{2}{K}\Delta R_{4,i}) n_i  - 1$ with $n_i = \tau_i-\tau_{i-1}$.}

The Bayesian simulator operates as follows:

\begin{itemize}
\item Sample $R=10$ initial break date vectors $\{\btau_i\}_{i=1}^{R}$ from the prior distribution.
\item Run an MCMC with $I$ iterations and at each iteration, for each $j=1,\ldots,R$, apply the D-DREAM Metropolis move:
\begin{enumerate}
\item Propose a new draw of the break parameter:
\begin{align*}%~\label{propDREAM}
\bz_j = \btau_j 
+ \text{round} \left[ \gamma(\delta, m) 
\left(\sum_{g=1}^{\delta} \btau_{r_1(g)} - \sum_{h=1}^{\delta} \btau_{r_2(h)} \right) + \xi \right] \,,
\end{align*}
with $\forall g, h = 1, 2, \ldots, \delta$ and $j \neq r_1(g)$, $r_2(h)$; $r_1(.)$ and $r_2(.)$ denote random integers uniformly distributed over support $[1,R]$. The $\text{round}[\cdot]$ operator picks the nearest integer and $\xi \sim \NORM(0,(0.0001) I)$. We set $\gamma(\delta,m) = \frac{2.38}{\sqrt{2\delta m}}$ and $\delta \sim \mathcal{U}(1,3)$.
\item Accept the proposal $\bz_j$ with probability:
\begin{align*}
\min \left\{\frac{f(\bz_j | \by_{1:T})}{f(\btau_j|\by_{1:T})}, 1 \right\} \,.
\end{align*}
\end{enumerate}
\end{itemize}

In practice, we set $I = 1,\!000$ and start collecting the draws after $\frac{I}{2}$ iterations. 

\end{document}